\documentclass[a4paper, amsfonts, amssymb, amsmath, reprint, showkeys, nofootinbib,superscriptaddress]{revtex4-2}
\usepackage[utf8]{inputenc}
\usepackage{amsmath,empheq}
\usepackage{braket}
\usepackage{graphicx}
\usepackage{textcomp}
\usepackage{gensymb}
\usepackage{hyperref}
\usepackage{mathtools}
\usepackage{dsfont}
\usepackage{bbold}

\newcommand{\RNum}[1]{\uppercase\expandafter{\romannumeral #1\relax}}

\begin{document}
\title{Prospects of cooling a mechanical resonator 
with a transmon qubit in c-QED setup}

\author{Sourav~Majumder}
\affiliation{Department of Physics, Indian Institute of Science, 
Bangalore-560012 (India)}

\author{Tanmoy~Bera}
\affiliation{Department of Physics, Indian Institute of Science, 
	Bangalore-560012 (India)}

\author{Vibhor~Singh}
% \email[]{v.singh@iisc.ac.in} 
\affiliation{Department of Physics, Indian Institute of Science, 
	Bangalore-560012 (India)}

\begin{abstract}

Hybrid devices based on the superconducting qubits have emerged as a 
promising platform for controlling the quantum states of macroscopic 
resonators. The nonlinearity added by a qubit can be a valuable 
resource for such control. Here we study a hybrid system consisting
of a mechanical resonator longitudinally coupled to a transmon qubit. 
The qubit readout can be done by coupling to a readout mode like in 
c-QED setup. The coupling between the mechanical resonator and transmon 
qubit can be implemented by modulation of the SQUID inductance. 
In such a tri-partite system, we analyze the steady-state occupation 
of the mechanical mode when all three modes are dispersively coupled. 
We use the quantum-noise and the Lindblad formalism to show that the sideband 
cooling of the mechanical mode to its ground state is achievable.
We further experimentally demonstrate that measurements of the
thermomechanical motion is possible in the dispersive limit, 
while maintaining a large coupling between qubit and mechanical mode.
Our theoretical calculations suggest that
single-photon strong coupling is within the experimental reach in 
such hybrid devices.
\end{abstract}

\date{\today}
\maketitle

\section{Introduction:}\label{intro}

Control over the quantum states of a mechanical resonator by 
coupling them to optical modes can have several potential 
applications in the field of quantum technologies \cite{barzanjeh_optomechanics_2022}. 
The traditional cavity-optomechanics based approach of coupling 
a mechanical resonator to an optical mode via the radiation-pressure 
interaction has been quite successful \cite{aspelmeyer_cavity_2014,teufel_sideband_2011,chan_laser_2011,wollman_quantum_2015,ockeloen-korppi_stabilized_2018, peterson_ultrastrong_2019,kotler_direct_2021, wollack_quantum_2022}.
While the radiation-pressure mediated coupling in such devices 
is nonlinear, its magnitude is usually small in most
applications.
Further, due to the dispersive interaction, the effects originating 
from the Kerr-term are strongly suppressed \cite{rabl_photon_2011,nunnenkamp_single-photon_2011}.
To mitigate the limitations of linear cavity optomechanics,  hybrid devices based on 
the strong nonlinearity of qubits have been proposed and developed
 \cite{rabl_cooling_2010,xiang_hybrid_2013, clerk_hybrid_2020}.  
These proposals explore their performance from the sideband cooling of the mechanical 
resonator \cite{martin_ground-state_2004} to the matter-interferometry \cite{khosla_displacemon_2018}, while considering a wide range of
two-level systems such as superconducting qubits \cite{martin_ground-state_2004, jaehne_ground-state_2008, hauss_dissipation_2008, wang_cooling_2009, nongthombam_ground-state_2021, wang_two-color_2018,manninen_enhancement_2022}, quantum-dots \cite{wilson-rae_laser_2004}, and nitrogen vacancy defects in diamond \cite{rabl_strong_2009}.
Particularly,
in the microwave domain, experimental realization of several hybrid 
devices have been shown using the nonlinearity of a superconducting qubit \cite{pirkkalainen_cavity_2015}, 
Josephson capacitance \cite{pirkkalainen_hybrid_2013,viennot_phonon-number-sensitive_2018},
Josephson inductance \cite{rodrigues_coupling_2019,schmidt_sideband-resolved_2020,zoepfl_single-photon_2020,bera_large_2021},
and piezo-electricity \cite{oconnell_quantum_2010,arrangoiz-arriola_resolving_2019}.
Among these different schemes, the electromechanical coupling
stems from charge or flux modulation, and its tunability is controlled 
by the external applied magnetic field. 
Recently,
the magnetic flux-mediated coupling approach have shown 
promising experimental results \cite{rodrigues_coupling_2019}.
These systems have demonstrated large electromechanical coupling \cite{schmidt_sideband-resolved_2020,zoepfl_single-photon_2020,bera_large_2021}, 
four-wave-cooling of the mechanical resonator to near the quantum ground 
state \cite{bothner_four-wave-cooling_2022}, and Lorentz-force induced 
backaction on the mechanical resonator \cite{luschmann_mechanical_2022}.
Motivated by the progress on flux-mediated 
approach, here we investigate a coupled three-mode system 
consisting of a mechanical mode, transmon qubit, and a readout cavity. 
From the practical point of view, the additional readout cavity 
is useful ingredient to consider as it allows the quantum non-demolishing 
(QND) measurement of qubit mode in circuit-QED setup \cite{gambetta_qubit-photon_2006,blais_circuit_2021}.
While a mechanical mode coupled to a two-level
system has been studied extensively in the past \cite{martin_ground-state_2004, zhang_cooling_2005, jaehne_ground-state_2008, rabl_cooling_2010, kounalakis_flux-mediated_2020}, 
the focus of our
investigation has been on treating the transmon qubit as a weakly
anharmonic oscillator. In addition, we theoretically and
experimentally address the readout of the mechanical mode 
when transmon is detuned far away from the readout cavity.
This regime is particularly important as large electromechanical 
coupling with the qubit mode can be achieved. 
Using the quantum-Langevin equation of motion \cite{gardiner_quantum_2004}, and Lindblad 
formalism \cite{lindblad_generators_1976}, we analyze the possibility of sideband cooling
of the mechanical resonator. 
Experimentally, we use a two-tone method to measure the 
thermo-mechanical motion, and compare it with analytical results. 
This paper is organised as follows: In part~\ref{model}, we discuss
the theoretical model of the three coupled modes. 
We solve the system's equations of motion in part~\ref{eq_motion}. 
The analytical solution of the system is analyzed in 
part~\ref{qubit_mec_psd}, where we have shown the possibility of cooling 
the mechanical resonator. In the part~\ref{expt}, 
we show experimental and analytical results discussing 
the detection of mechanical motion in the dispersive regime of the 
cavity and the qubit mode. We summarize and conclude our discussion in part~\ref{conclusion}.

\section{Theoretical model:}\label{model}

%%%%%%%%%%%%%%%%%%% FIG1 description %%%%%%%%%%%%%%%%%%
\begin{figure}
\begin{center}
\includegraphics[width = 85 mm]{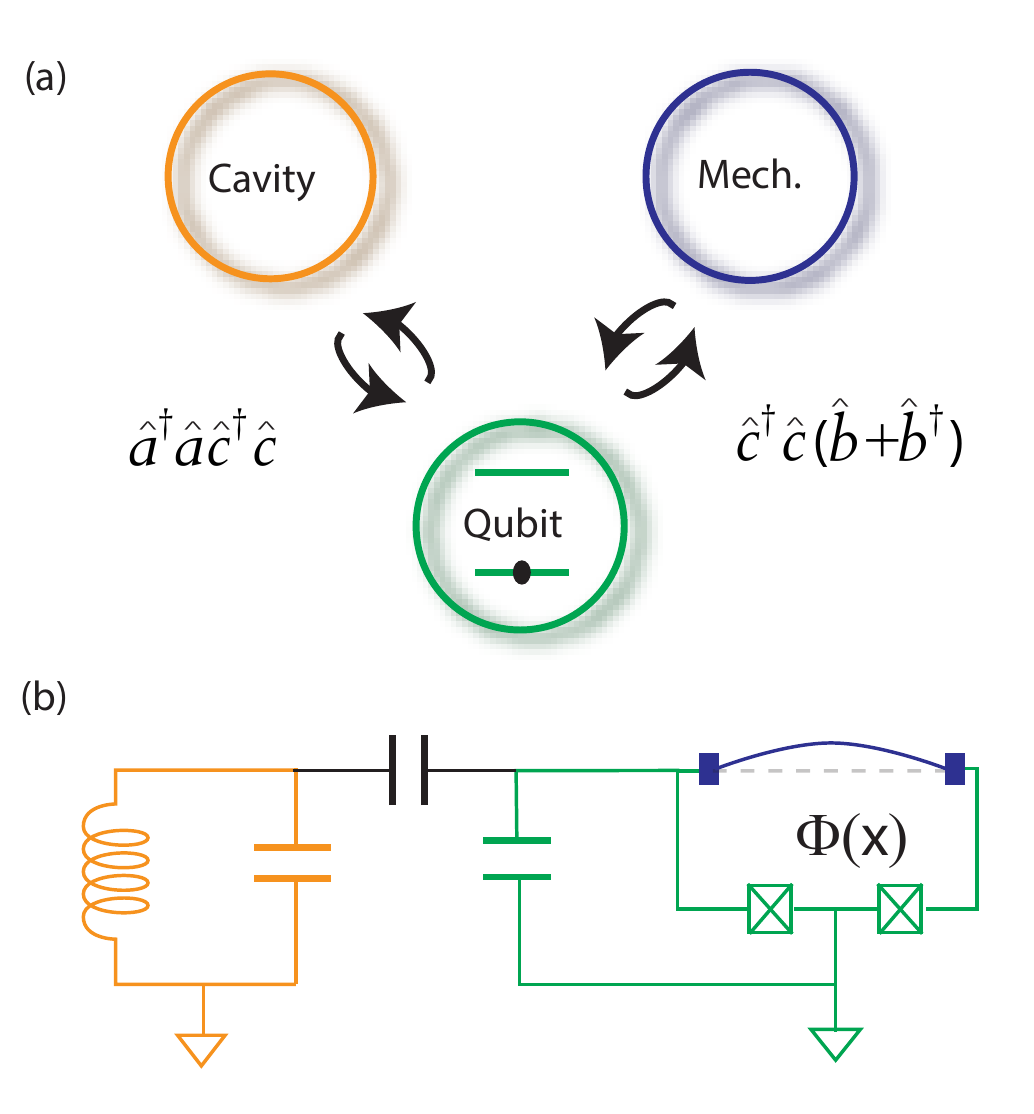}
\caption{(a) A conceptual schematic of the three-mode hybrid device 
showing a linear cavity coupled to a qubit which in turn couples to 
a mechanical resonator. A direct coupling 
between the cavity and the mechanical mode is not considered. 
(b) A possible implementation using a frequency tunable transmon 
qubit, where coupling to mechanical mode is achieved by embedding
it the SQUID loop and by applying a constant magnetic field. A magnetic 
field perpendicular to the SQUID loop couples the in-plane mechanical 
mode to the qubit, while parallel magnetic field couples the qubit to 
the out-of-plane mechanical mode. }
\label{fig1}
\end{center}
\end{figure}

We consider a coupled system where the mechanical mode modulates 
the transmon qubit frequency, therefore resulting in a longitudinal coupling.
Such coupling between transmon qubit and the mechanical resonator 
can be implemented by embedding a mechanical resonator into 
the SQUID loop of the qubit.
In addition, the qubit couples to a linear mode (the readout cavity)
transversely as in the circuit-QED setup. 
A schematic diagram of the system and a possible implementation with 
the equivalent circuit diagram are shown in the Fig.~\ref{fig1} (a) and (b).

Using the dispersive approximation
between the transmon and the readout cavity, we arrive at the following
system Hamiltonian:

\begin{multline}
\hat{\mathcal{H}}_0 = \omega_c \hat{a}^\dagger\hat{a} + \omega_q \hat{c}^\dagger\hat{c} 
- \frac{\alpha_q}{2} \hat{c}^\dagger\hat{c}^\dagger\hat{c}\hat{c} +
\omega_m \hat{b}^\dagger\hat{b} \\
+ \chi~\hat{a}^\dagger\hat{a}\hat{c}^\dagger\hat{c} +
g_0~\hat{c}^\dagger\hat{c}(\hat{b} + \hat{b}^\dagger),
\label{dispersive hamiltonian}
\end{multline}
where $\hat{a}$($\hat{a}^\dagger$), $\hat{c}$($\hat{c}^\dagger$), 
$\hat{b}$($\hat{b}^\dagger$) are the annihilation(creation) 
operators for the cavity, qubit and the mechanical mode of 
frequency $\omega_c$, $\omega_q$, $\omega_m$, respectively. 
The Kerr-nonlinearity of the transmon is denoted as $\alpha_q$. 
The last two terms are the interaction terms between the modes, 
where the dispersive coupling between the qubit 
and the cavity is $\chi$. The radiation-pressure type 
coupling between the transmon and the mechanical mode is denoted 
by the single photon coupling rate $g_0$.

Two additional drive terms of amplitude $\delta$ and 
$\epsilon$ at frequency of $\omega_L$ (near $\omega_c$) 
and $\omega_d$ (near $\omega_q$) are added to the 
Hamiltonian. We can write the drive Hamiltonian as,
\begin{equation}
\hat{\mathcal{H}}_{d} = \delta~(\hat{a}~e^{+{i} \omega_{L}t} + \hat{a}^\dagger~e^{-{i} \omega_{L}t})+
\epsilon~(\hat{c}~e^{+{i} \omega_{d}t} + \hat{c}^\dagger~e^{-{i} \omega_{d}t}).
\end{equation}

By carrying out rotating frame transformations, given by the unitary 
operators $U^a = \exp{\left[{i}\omega_{L}\hat{a}^\dagger\hat{a}t\right]}$
and $U^c = \exp{\left[{i}\omega_{d}\hat{c}^\dagger\hat{c}t\right]}$, 
the transformed Hamiltonian can be written as, 

\begin{multline}
\hat{\mathcal{H}} = -\Delta_c \hat{a}^\dagger\hat{a} - \Delta_q \hat{c}^\dagger\hat{c} - \frac{\alpha_q}{2} \hat{c}^\dagger\hat{c}^\dagger\hat{c}\hat{c} +
\omega_m \hat{b}^\dagger\hat{b} +
\chi~\hat{a}^\dagger\hat{a}\hat{c}^\dagger\hat{c} \\+
g_0~\hat{c}^\dagger\hat{c}(\hat{b} + \hat{b}^\dagger) + 
\delta~(\hat{a} + \hat{a}^\dagger)+
\epsilon~(\hat{c} + \hat{c}^\dagger),
\label{rotated hamiltonian}
\end{multline}
where $\Delta_c = \omega_L - \omega_c$ and $\Delta_q = \omega_d - \omega_q$.
The transformed Hamiltonian is time-independent in this frame of rotation.
For further analysis, we shift the frame to mean field using 
the following displacement transformation,
\begin{equation}
    \mathcal{D}(\alpha, \mu, \beta) = \exp{\left[\alpha(\hat{a} - \hat{a}^\dagger) + \mu(\hat{c} - \hat{c}^\dagger) + \beta(\hat{b} - \hat{b}^\dagger)\right]},
\end{equation}
where $\alpha$, $\mu$, $\beta$ are real scalar quantities. 
For a particular choice of $\alpha = \Bar{\alpha}$, $\mu = \Bar{\mu}$ 
and $\beta = \Bar{\beta}$, all the drive terms (terms proportional 
to $\hat{a} + \hat{a}^\dagger$, $\hat{b} + \hat{b}^\dagger$, and $\hat{c} + \hat{c}^\dagger$) 
get cancelled.
After dropping the third and higher order terms, we arrive at 
the following effective Hamiltonian,

\begin{multline}
    \hat{\mathcal{H}^\prime} \approx -\tilde{\Delta}_c\hat{a}^\dagger\hat{a} -
    \tilde{\Delta}_q\hat{c}^\dagger\hat{c} - \eta(\hat{c}^2 + \hat{c}^{\dagger 2}) + \omega_m \hat{b}^\dagger\hat{b} \\+ J(\hat{a} + \hat{a}^\dagger)(\hat{c} + \hat{c}^\dagger) + g(\hat{c} + \hat{c}^\dagger)(\hat{b} + \hat{b}^\dagger),
    \label{final hamiltonian}
\end{multline}

where $\tilde{\Delta}_c = \Delta_c - \chi \Bar{\mu}^2$, $\tilde{\Delta}_q = \Delta_q + 2\alpha_q\Bar{\mu}^2 - \chi\Bar{\alpha}^2 - 2g_0\Bar{\beta}$, $\eta = \frac{\alpha_q\Bar{\mu}^2}{2}$, $J = \chi\Bar{\alpha}\Bar{\mu}$ and $g = g_0\Bar{\mu}$.
It might be important to underline here that the coupling rates
$g$ and $J$ as defined above are the scaled coupling rates.
They show the scaling with drive tone amplitude similar to the case in linear optomechanical device.

\section{Equations of motion:} \label{eq_motion}
Dynamics of the system depends on various decay rates associated 
with different modes and drive amplitudes. We write the equations 
of motion for the field operators while incorporating all the noise 
operators and decay rates as, %
\begin{subequations}
\begin{align}
    \dot {\hat{a}} &= - {i}\left[ \hat{a}, \hat{\mathcal{H}^\prime} \right] - \frac{\kappa}{2} \hat{a} + \sqrt{\kappa_{ex}}~\hat{a}_{in} + \sqrt{\kappa_{0}}~\hat{f}_{in}, \\
    \dot {\hat{c}} &= - {i}\left[ \hat{c}, \hat{\mathcal{H}^\prime} \right] - \frac{\Gamma}{2} \hat{c} + \sqrt{\Gamma_{ex}}~\hat{c}_{in} + \sqrt{\Gamma_0}~\hat{\xi}_{in}, \\
    \dot {\hat{b}} &= - {i}\left[ \hat{b}, \hat{\mathcal{H}^\prime} \right] - \frac{\gamma_m}{2} \hat{b} + \sqrt{\gamma_m}~\hat{b}_{in},
\end{align}\label{eom}
\end{subequations}
where $\hat{a}_{in}$, $\hat{c}_{in}$, $\hat{b}_{in}$, $\hat{f}_{in}$, 
$\hat{\xi}_{in}$ are noise operators of cavity, qubit and mechanical mode, respectively.
The mechanical energy dissipation rate is $\gamma_m$.
The internal, external and total cavity (qubit) dissipation rates are $\kappa_0$ ($\Gamma_0$), $\kappa_{ex}$ ($\Gamma_{ex}$), and $\kappa$ ($\Gamma$), respectively.

This set of equations can be easily solved by performing a Fourier transformation, defined as $x[\omega] = \mathcal{F}\left[x(t)\right] = \int_{-\infty}^{+\infty} x(t) \mathrm{e}^{{i}\omega t} \mathrm{d}t$, of the equations. 
We now define a field vector $u[\omega] = \begin{bmatrix}
                \hat{a}[\omega] & (\hat{a}^{\dagger})[\omega] & \hat{c}[\omega] & (\hat{c}^{\dagger})[\omega] & \hat{b}[\omega] & (\hat{b}^{\dagger})[\omega]
            \end{bmatrix}^T$ and evaluate its governing equation of the form,
\begin{equation}
    u[\omega] = (-{i}\omega \mathbb{1} - A)^{-1}~r[\omega] = \mathcal{B}~r[\omega],
    \label{solution}
\end{equation}
where,
\begin{equation}
    r[\omega] = \begin{bmatrix}
                \sqrt{\kappa_{ex}}~\hat{a}_{in}[\omega] + \sqrt{\kappa_{0}}~\hat{f}_{in}[\omega] \\
                \sqrt{\kappa_{ex}}~(\hat{a}_{in}^{\dagger})[\omega] + \sqrt{\kappa_{0}}~(\hat{f}_{in}^{\dagger})[\omega] \\
                \sqrt{\Gamma_{ex}}~\hat{c}_{in}[\omega] + \sqrt{\Gamma_{0}}~\hat{\xi}_{in}[\omega] \\
                \sqrt{\Gamma_{ex}}~(\hat{c}_{in}^{\dagger})[\omega] + \sqrt{\Gamma_{0}}~(\hat{\xi}_{in}^{\dagger})[\omega] \\
                \sqrt{\gamma_m}~\hat{b}_{in}[\omega] \\
                \sqrt{\gamma_m}~(\hat{b}_{in}^{\dagger})[\omega]
            \end{bmatrix}
\end{equation}
The matrix $\mathcal{B}$ can be calculated from Eq.~(\ref{final hamiltonian}) 
and Eq.~(\ref{eom}), as
\begin{equation}
    \mathcal{B} = \begin{bmatrix}
            1/\chi_c & 0 & {i}J & {i}J & 0 & 0 \\
            0 & 1/\tilde{\chi}_c & -{i}J & -{i}J & 0 & 0 \\
            {i}J & {i}J & 1/\chi_q & -2{i}\eta & {i}g & {i}g \\
            -{i}J & -{i}J & 2{i}\eta & 1/\tilde{\chi}_q & -{i}g & -{i}g \\
            0 & 0 & {i}g & {i}g & 1/\chi_m & 0 \\
            0 & 0 & -{i}g & -{i}g & 0 & 1/\tilde{\chi}_m
    \end{bmatrix}^{\scalebox{1}{-1}}.
\end{equation}
All $\chi$'s in the matrix represent the susceptibility 
of the modes, defined as,
\begin{equation*}
    \chi_c[\omega] = \frac{1}{- {i}\omega - {i}\tilde{\Delta}_c + \frac{\kappa}{2}};~  \tilde{\chi}_c[\omega] = \frac{1}{- {i}\omega + {i}\tilde{\Delta}_c + \frac{\kappa}{2}}
\end{equation*}
\begin{equation*}
    \chi_q[\omega] = \frac{1}{- {i}\omega - {i}\tilde{\Delta}_q + \frac{\Gamma}{2}};~ 
    \tilde{\chi}_q[\omega] = \frac{1}{- {i}\omega + {i}\tilde{\Delta}_q + \frac{\Gamma}{2}}
\end{equation*}
\begin{equation*}
    \chi_m[\omega] = \frac{1}{- {i}\omega + {i}\omega_m + \frac{\gamma_m}{2}};~ 
    \tilde{\chi}_m[\omega] = \frac{1}{- {i}\omega - {i}\omega_m + \frac{\gamma_m}{2}}.
\end{equation*}
From Eq.~\ref{solution}, we can solve for the field operators. 
Further, we define the spectrum of any mode as,
\begin{equation}
    S_{\mathcal{O}}(\omega) = 
    \frac{1}{2\pi} \int_{-\infty}^{+\infty} \langle (\hat{\mathcal{O}}[\omega^\prime])^\dagger\hat{\mathcal{O}}[\omega]\rangle \mathrm{d}\omega^\prime.
    \label{spectrum equation}
\end{equation}

Eq.~\ref{spectrum equation} and the solution of field operators can be 
used to get the spectrum of the modes. The detailed calculations and the 
correlators of noise operators are given in Appendix~\textbf{A}. The
calculated spectrum as follows,

\begin{multline}
    S_{\text{x}}(\omega)\Big|_{\text{x}\epsilon \{1,3,5\}} =
    n_m^i \gamma_m(|\mathcal{B}_{\text{x}5}[\omega]|^2 + |\mathcal{B}_{\text{x}6}[\omega]|^2) + \\
    \kappa |\mathcal{B}_{\text{x}2}[\omega]|^2 + 
    \Gamma |\mathcal{B}_{\text{x}4}[\omega]|^2 + 
    \gamma_m|\mathcal{B}_{\text{x}6}[\omega]|^2,
    \label{spectrum}
\end{multline}
where $n_m^i$ is the initial phonon occupation in the mechanical mode. 
The indexing $\{S_{1}, S_{3}, S_{5}\}$ maps to the spectrum of cavity, qubit 
and mechanics as $\{S_{a}, S_{c}, S_{b}\}$, respectively.

\section{Spectrum of the qubit and the mechanical mode}\label{qubit_mec_psd}

%%%%%%%%%%%%%%%%%%% FIG2 description %%%%%%%%%%%%%%%%%
\begin{figure}[htb]
\begin{center}
\includegraphics[width = 85 mm]{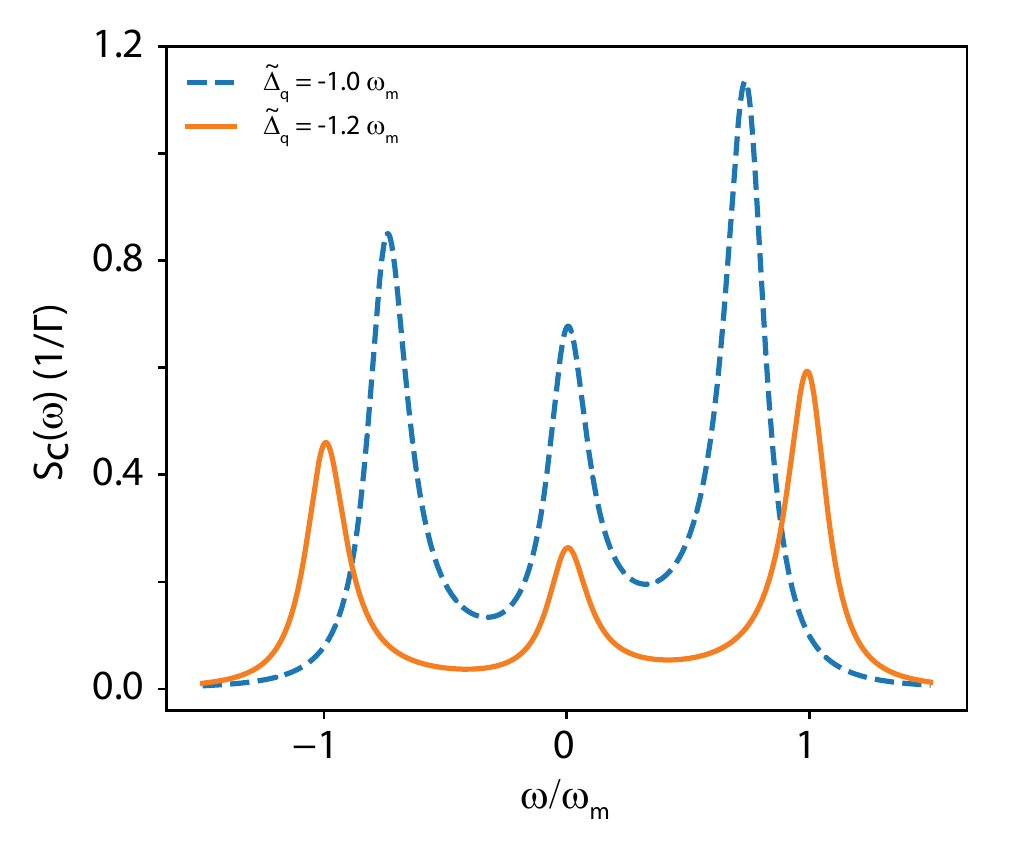}
\caption{Plot of the qubit spectrum	for two different 
values of drive detunings, $\tilde{\Delta}_q = - 1.0~\omega_m$ and $\tilde{\Delta}_q = - 1.2~\omega_m$.
The parameters used for the plots: $\tilde{\Delta}_c = 0$, $\omega_m = 2\pi\times6$ MHz, $J = 2\pi\times0.8$ MHz, $g = 2\pi\times2$ kHz, $\kappa = 2\pi\times4$ MHz, $\omega_m/\Gamma = 5$, $\gamma = 2\pi\times6$ Hz, and $\eta = 2\pi\times2$ MHz.
}
\label{fig2}
\end{center}
\end{figure}

%%%%%%%%%%%%%%%%%%% FIG3 description %%%%%%%%%%%%%%%%%%%%%%%%
\begin{figure*}[htb]
\begin{center}
\includegraphics[width = 150 mm]{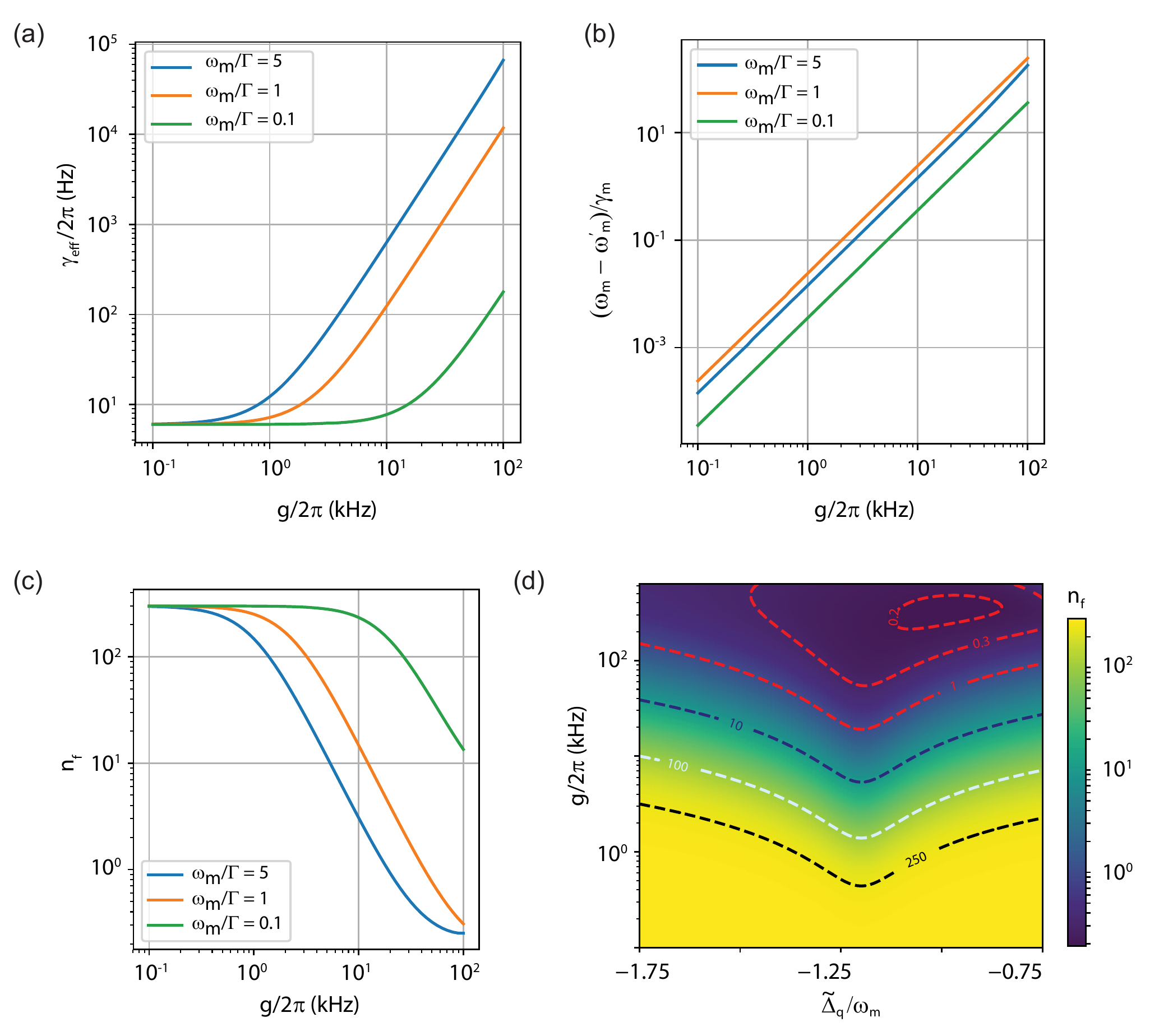}
\caption{\textbf{Cooling of the mechanical mode: }
The spectrum of the mechanical mode is analyzed to characterize the
effect of back action arising from the drive tone near the qubit 
frequency $\omega_q$. The extracted parameters for effective 
mechanical linewidth and shift in the mechanical resonant frequency 
as the electromechanical coupling between the qubit and the mechanical 
mode is varied, are shown in (a) and (b). Panel (c) shows the
final phonon occupancy of the mechanical mode. It is extracted by calculating 
the area under the Lorentzian in mechanical spectrum. 
For large qubit-mechanics coupling a final phonon occupation 
well-below 1 can be achieved for various sideband parameters. 
(d) Final phonon occupancy as a function of qubit-mechanics coupling 
and scaled detuning between the drive and the qubit frequency for 
$\omega_m/\Gamma = 5$. The parameters used for the plots are: 
$\tilde{\Delta}_c = 0$, $\omega_m = 2\pi\times6$~MHz, 
$J = 2\pi\times0.8$~MHz, $\eta = 2\pi\times2$~MHz, 
$\kappa = 2\pi\times4$~MHz, $\gamma = 2\pi\times6$~Hz, $n_m^i = 300$. 
For the plot in panel (a), (b), and (c), we use $\tilde{\Delta}_q = -1.2~\omega_m$ 
as the detuning.
}
\label{fig3}
\end{center}
\end{figure*}

In this section, we discuss the best cooling scenario of the mechanical resonator by inspecting the qubit spectrum.
%
% To understand cooling, we compute the qubit spectrum when a small 
% drive signal near the qubit frequency is added. 
% %
Fig.~\ref{fig2} shows the spectrum of the transmon qubit for two different
detuning of the drive tone,($\tilde{\Delta}_q = - 1.0~\omega_m$ and $\tilde{\Delta}_q = - 1.2~\omega_m$).
In presence of a the nearly red detuned drive on qubit mode,
its spectrum becomes asymmetric. The cooling rate is 
calculated from the asymmetry of the spectrum, which is large 
for a specific drive position.
In the weak coupling regime ($g << \Gamma$), 
the cooling rate for the mechanical resonator is given by
$\Gamma_c = 2[g^2 (S_{c}(\omega_m) - S_{c}(-\omega_m)) + \gamma_m]$ \cite{jaehne_ground-state_2008, rabl_cooling_2010}.
The optimum cooling rate, as seen from Fig.~\ref{fig2}, is a 
function of the position of the drive \cite{jaehne_ground-state_2008}.
Unlike a linear cavity as a bath for cooling,
the cooling rate of a mechanical resonator for an anharmonic 
oscillator (the qubit) depends on the position of the cooling 
tone applied and the anharmonicity of the resonator mode. 
This is a direct consequence of the Kerr-term. In the steady
state, the final phonon occupancy
can be calculated from the cooling rate and the qubit spectrum as, 
\begin{equation}
    n_f = 2\frac{n_m^i \gamma_m}{\Gamma_c} + 2 g^2 \frac{S_{c}(-\omega_m)}{\Gamma_c}.
\end{equation}
To further understand the backaction on the mechanical resonator due 
to a drive on the qubit mode, we compute the mechanical 
spectrum $S_{b}(\omega)$. 
In the steady state, the mean phonon occupancy of the mechanical 
mode can be calculated as $n_f = \frac{1}{2\pi}\int S_{b}(\omega) d\omega$, 
which is the area under the Lorentzian in the mechanical mode spectrum. 
While it is possible to reduce the expression of the 
mechanical spectrum to a Lorentzian form, we find 
it more efficient to compute the spectrum and carry out a numerical 
fit to extract the effective linewidth and the effective resonant
frequency. 
Fig.~\ref{fig3}(a)  and Fig.~\ref{fig3}(b) show the linewidth 
broadening and resonant frequency shift of the mechanical mode,
for a red detuned ($\tilde{\Delta}_q = -1.2~\omega_m$) qubit drive.
The back-action on the mechanical resonator from the drive on 
qubit is reflected in the change of mechanical frequency 
and an increase in the effective linewidth.
The final phonon occupation is plotted in Fig.~\ref{fig3}(c)
for different value of sideband parameter $\omega_m/\Gamma$. 
It is evident from the figure that in the steady driving of the qubit, the 
final phonon occupancy strongly depends on sideband parameter $\omega_m/\Gamma$. 
A larger value of sideband parameter offers better cooling of the 
mechanical mode.
It is important to underline here that the cooling to the quantum 
ground state of the mechanical resonator is possible well before
entering the strong coupling regime, $g~\gtrsim \text{max}~(\Gamma, \kappa)$. 
To gain insight into the spectrum calculation, we consider 
a simpler case when qubit anharmonicity is set to zero 
$\eta = 0$, and it is being driven at the lower mechanical 
sideband $\tilde{\Delta}_q = - \omega_m$.
With these parameters and Eq.~\ref{spectrum equation},
the mechanical spectrum can be approximately written as,
\begin{equation}
S_{b}(\omega) = \frac{n_m^i \gamma_m \Gamma^2/(\Gamma^2 - 8 g^2)}{(\omega - \omega_m)^2 + \frac{(4 g^2 + \gamma_m \Gamma)^2}{4(\Gamma^2 - 8 g^2)}}.
\end{equation}

From this simplified expression of the mechanical spectrum,
we can write the effective line-width of the mechanical resonator as,
$\gamma_{eff} = \frac{4 g^2 + \gamma_m \Gamma}{\sqrt{\Gamma^2 - 8 g^2}} 
\simeq \gamma_m(C+1)$, where $C = \frac{4g^2}{\gamma_m \Gamma}$ is defined
as the cooperativity. 
Similarly, the final mean phonon occupation can be written as,
$n_f = \frac{n_m^i \gamma_m \Gamma^2}{4g^2 + \gamma_m \Gamma} \frac{1}{\sqrt{\Gamma^2 - 8g^2}} \simeq \frac{n_m^i}{1 + C}$ for $\Gamma \gg g$.
We note that in the limit of zero anharmonicity and weak coupling,
the results are consistent with that obtain from linear cavity optomechanics \cite{aspelmeyer_cavity_2014}. 

For the model Hamiltonian given by Eq.~\ref{final hamiltonian},
the mean phonon-occupation can 
also be obtained by solving Lindblad master equation.
Here, we obtain the equations of motion for the expectation 
values of mode operators and solve for the 
steady-state solutions.
From this formalism, we calculate the steady-state occupancy 
in the mechanical mode for the various drive detuning $\tilde{\Delta}_q$
and coupling $g$.
Fig.~\ref{fig3}(d) shows the color plot of the final phonon 
occupation for the sideband parameter of $\omega_m/\Gamma = 5$.
We can see that the optimum cooling can be achieved near 
the detuning of $\tilde{\Delta}_q \approx-1.2 \omega_m$.
It is important to emphasize here that the lowest phonon
occupation of the mechanical resonator depends on the device
parameters, such as qubit thermal occupation and dissipation rate
$\Gamma$. 
For the calculations presented in this section, 
we assumed the thermal occupation of the qubit and readout cavity 
to be zero.
Another important parameter that affects the ultimate performance 
of the sideband cooling is sideband parameter $\omega_m/\Gamma$ \cite{rabl_cooling_2010}, and cooling to the ground state can only 
be achieved in sideband-resolved limit $\omega_m/\Gamma~\gtrsim~1$.

\section{Experimental Details}\label{expt}

%%%%%%%%%%%%%%%%%%% FIG4 description %%%%%%%%%%%%%%%%%%
\begin{figure}
\begin{center}
\includegraphics[width = 85 mm]{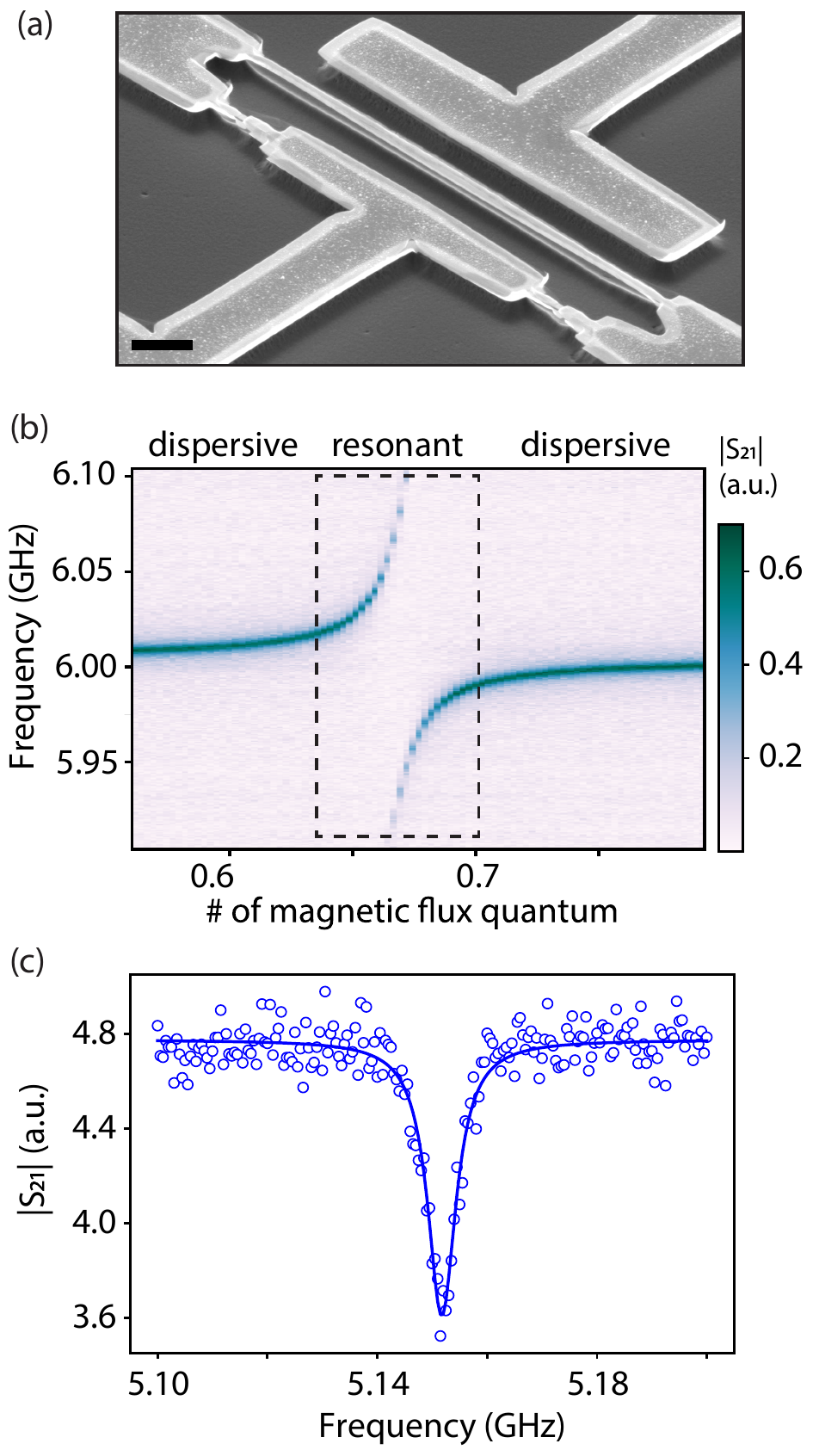}
\caption{
(a) A SEM-image of the device showing the suspended part of the SQUID 
loop and the Josephson junctions. The length and width of the 
nanowire is 40~$\mu$m and 200~nm, respectively. 
The scale bar corresponds to 5~$\mu$m.
(b) Color plot of the cavity transmission $|S_{21}|$ as a 
function of the magnetic flux through the SQUID loop. 
(c) Two-tone measurements spectroscopic linewidth of the qubit 
in the dispersive regime. 
}
\label{fig4}
\end{center}
\end{figure}

After discussing the performance of the sideband cooling when 
the qubit is dispersively coupled to the readout cavity, we address
the next question on the possibility of the mechanical readout. 
In the dispersive regime, there is no direct coupling between 
the cavity and the mechanical resonator. The modulation of qubit 
frequency translates to the cavity mode via dispersive coupling,
and thus creating an effective coupling between the cavity and the 
mechanical motion. 
By tuning the transmon qubit frequency near half flux quantum, 
a large electromechanical coupling with the qubit mode can
be obtained. However, when $|\omega_q-\omega_c|$ is large,
the effective coupling between the cavity and mechanical
mode is suppressed.
Next, we show that the addition of cooling tone near the qubit 
frequency is helpful for the readout of the mechanical motion.

%%%%%%%%%%%%%%%%%%% FIG5 description %%%%%%%%%%%%%%%%%%%%%%
\begin{figure*}[htb]
\begin{center}
\includegraphics[width = 165 mm]{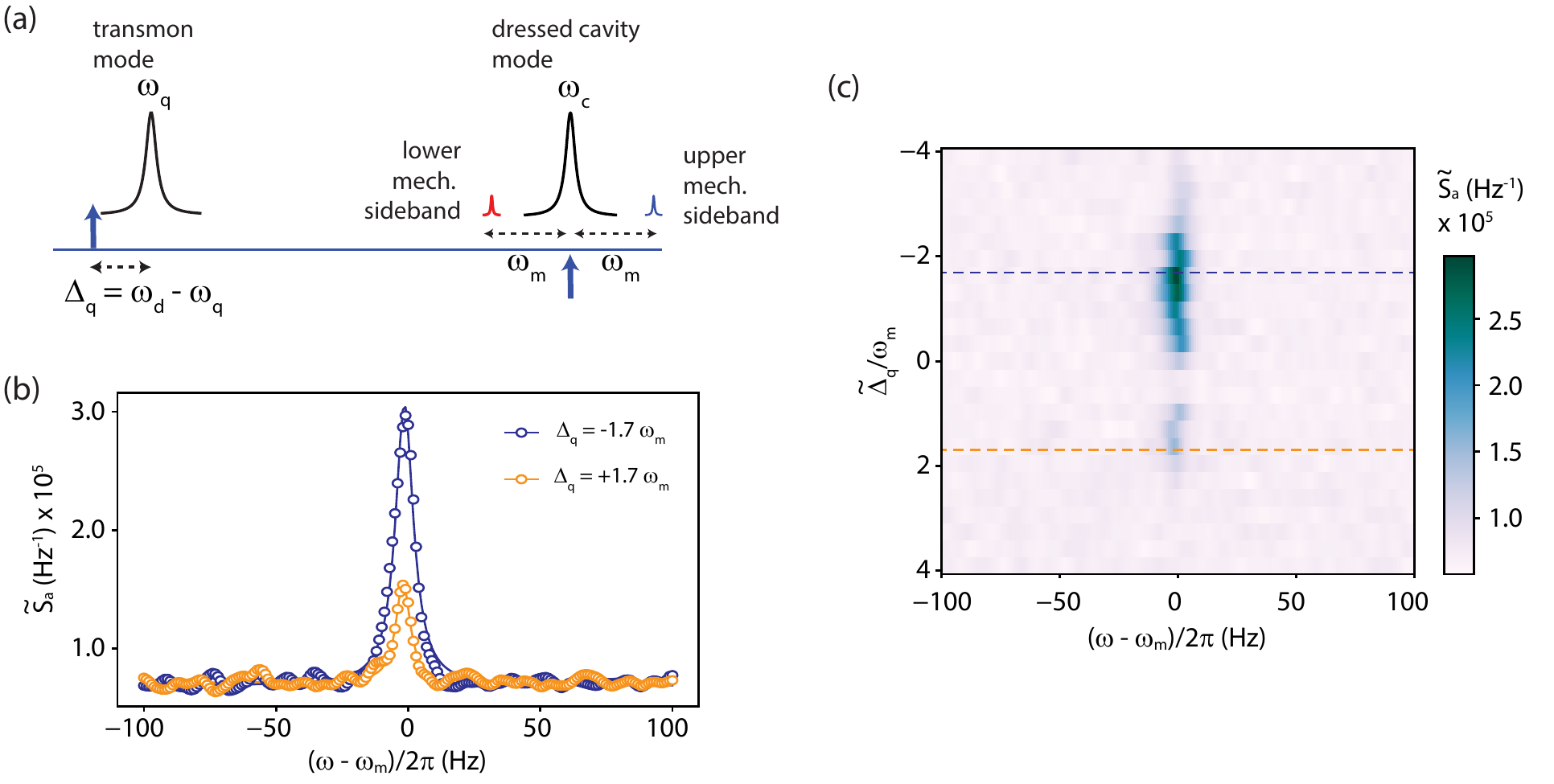}
\caption{\textbf{Experimental Data: }
Power spectral density of the cavity mode is measured while varying 
the drive detuning from the qubit mode. (a) Schematic of the 
measurement process. A drive is present near the qubit 
mode. The detuning between the qubit and the drive frequency is 
being changed in the measurement. A probe of frequency $\omega_c$
is added and its lower and upper mechanical sidebands
are recorded with a spectrum analyzer.
(b) The spectral density is shown for the drive detuning of 
$\tilde{\Delta}_q = -1.7~\omega_m$ and $\tilde{\Delta}_q = +1.7~\omega_m$. 
We can observe the difference in spectral height as the 
detuning change sign. The mechanical resonator has a frequency 
of $\omega_m/2\pi\approx$~5.9~MHz and a linewidth $\gamma_m/2\pi=$~6~Hz.
(c) A colorplot of normalized spectral density as a function of 
detuning and measurement frequency.
}
\label{fig5}
\end{center}
\end{figure*}

% %%%%%%%%%%%%%%%%%%% FIG6 description %%%%%%%%%%%%%%%%%

\begin{figure}
\begin{center}
\includegraphics[width = 85 mm]{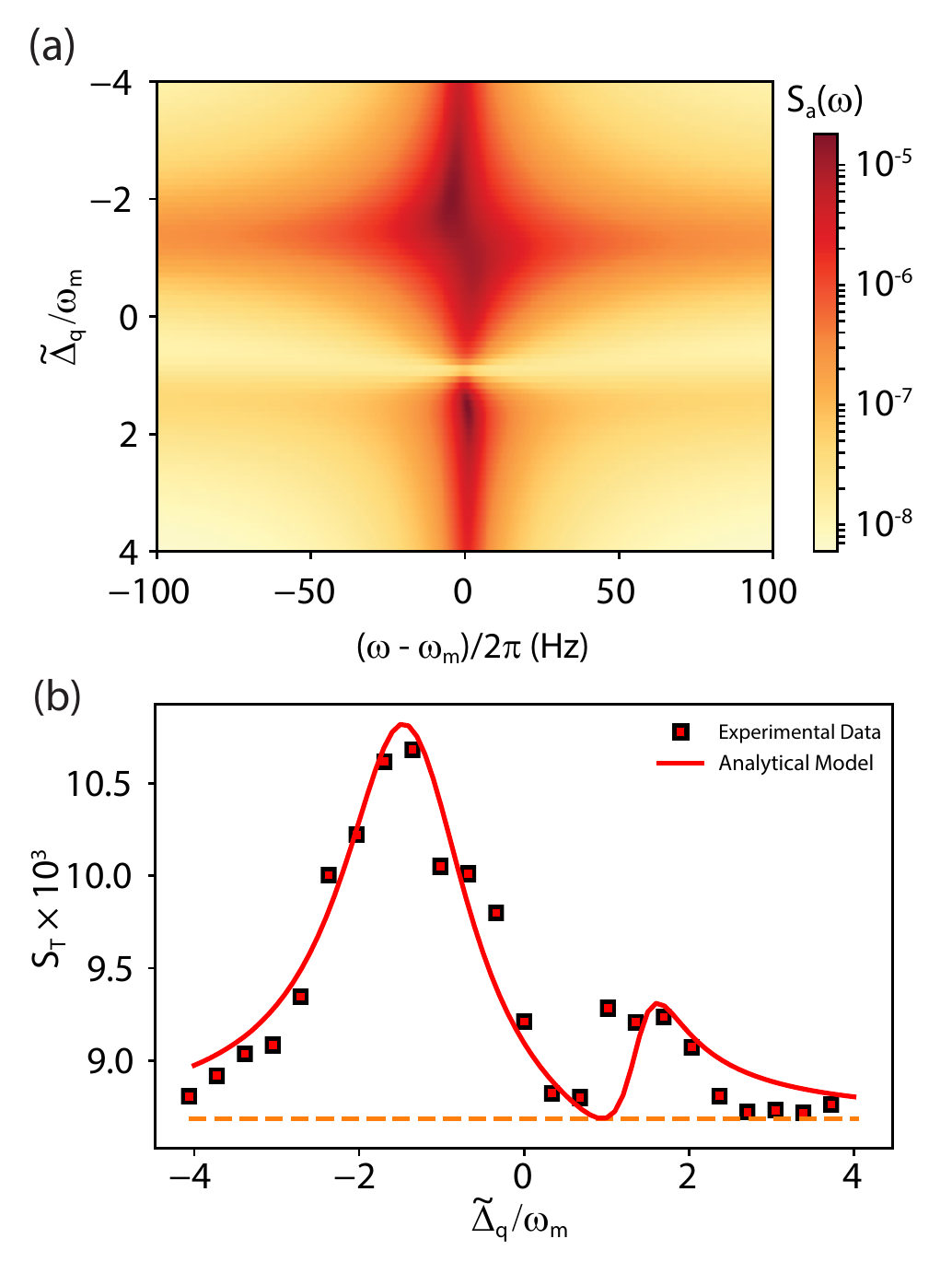}
\caption{
(a) We have evaluated the expression for the cavity mode 
spectrum from the theoretical model as a function of 
detuning $\tilde{\Delta}_q$  and frequency. 
Parameters are taken from the device studied here.
(b) Plot of integrated spectrum $S_T =\int S_{a}(\omega) d\omega$ for different detuning is calculated from the 
theoretical and experimental results. The square points 
indicate the experimental data, plotted as a function of drive
detuning ($\tilde{\Delta}_q$).
The solid curve is plotted for estimated device parameters 
from analytical expression. The dashed straight line indicates 
noise level of the measurements. The parameters used for the 
plots: $\tilde{\Delta}_c = 0$, $\omega_m = 2\pi\times 5.9$~MHz, 
$J = 2\pi\times 5.6$~MHz, $g = 2\pi\times 3.6$~kHz, 
$\eta = 2\pi\times 2.8$~MHz, $\kappa = 2\pi\times 4$~MHz, 
$\Gamma = 2\pi\times 8$~MHz, $\gamma_m = 2\pi\times 6$~Hz, 
$n_m^i$ = 350.
}
\label{fig6}
\end{center}
\end{figure}

For experimental realization, we use a device consisting 
of a transmon qubit with a doubly clamped suspended nanowire
embedded in the SQUID loop. For the qubit readout, we use a 3D copper 
rectangular waveguide cavity.
The scanning electron microscope (SEM) image of the device is shown
in Fig.~\ref{fig4}(a).  
The transmon, fabricated on a silicon substrate coated with highly stressed SiN, 
is designed to have tunable frequency realized via SQUID. One arm of the SQUID 
is made suspended to form a nanowire, essentially establishing the mechanical mode. 
The silicon substrate is placed inside the readout cavity and cool down to 20~mK in
a dilution refrigerator. A detailed description of the device fabrication methods and 
the measurement setup can be found in Ref.~\cite{bera_large_2021}.

Fig.~\ref{fig4}(b) shows the cavity transmission amplitude $|S_{21}|$ as the 
magnetic flux through the SQUID loop is varied.
When the qubit is brought in resonance with the cavity mode, the vacuum-Rabi 
splitting is observed and two hybrid modes emerge as indicated by the dashed 
box in Fig.~\ref{fig4}(b). 
From the avoided-crossing, we determine the qubit-cavity coupling 
strength to be $75$~MHz.
We measure the dressed cavity frequency to be 6.006~GHz, the 
maximum qubit frequency to be 7.8~GHz, and the qubit anharmonicity 
to be $-130$~MHz. We apply a magnetic field of $B\approx$~1.1~mT, perpendicular to
the plane of the SQUID loop. It couples the in-plane motion of the
mechanical resonator to the qubit.

To operate in the dispersive limit, we choose a qubit detuning
$\Delta = \omega_q - \omega_c$ of $-2\pi\times$900~MHz. 
A representative two-tone measurement of the qubit is shown in 
Fig.~\ref{fig4}(c). To record the mechanical motion at this
operating point, we apply two tones to the device, a drive tone 
near the qubit frequency and a probe tone near $\omega_c$ and 
record the mechanical sidebands of the probe tone using a spectrum 
analyzer.
The positioning of various frequencies and drive tones are shown in 
Fig.~\ref{fig5}(a).

\par

Fig.~\ref{fig5}(b) shows the recorded spectrum for two different detunings.
The experimentally measured microwave spectrum $S_{mw}(\omega)$ is normalized and 
represented in the units of intra-cavity photons defined 
as, $\tilde{S}_{a} = S_{mw}(\omega)/(\hbar\omega_c G \kappa_{ex} RBW )$, 
where $G$ is the estimated net gain of the output line, $\kappa_{ex}$ is the external 
coupling rate of the output port of the cavity, and $RBW$ is the resolution 
bandwidth of the spectrum analyzer.
Clearly, the spectrum has a larger peak for negative detuning
as compared to the one for the positive detuning.
This asymmetry becomes quite evident as the detuning of qubit drive is varied. 
Fig.~\ref{fig5}(c) shows the colorplot of $\tilde{S}_{a}$ as drive frequency
is varied across the qubit transition.
The mechanical resonator has a frequency of $\omega_m/2\pi\approx$~5.9~MHz
and a linewidth of $\gamma_m/2\pi\approx$~6~Hz. Here, we do not observe any 
backaction on the mechanical resonator. Both, the mechanical frequency 
and linewidth do not show any measurable change as the detuning
$\tilde{\Delta}_q$ is varied across the qubit frequency.
This is expected behavior within the experimental parameters. 
For these measurements, we estimated a single-photon coupling rate of
$g_0/2\pi\approx$~7.5~kHz, and measured a qubit linewidth of 
$\Gamma/2\pi\approx$~15~MHz. The lower sideband parameter
and single-photon coupling rate reduces the effect of back-action
from the qubit drive. 
Another aspect of the measurement is the the enhancement of the 
transduction and asymmetry with respect to $\tilde{\Delta}_q$.
Qualitatively, it can be understood from the qubit-cavity 
dispersive coupling and the Kerr-term of the qubit mode. A drive tone
near the qubit frequency acts like a parametric pump due to the 
qubit-nonlinearity,
resulting in the amplification of the field fluctuations due to electromechanical
coupling. Further, due to the dispersive interaction between the qubit and the
cavity mode, these field fluctuations result in the modulation of the 
intracavity probe field, and hence in an improved 
transduction. The asymmetry in the response is a direct manifestation 
of the weak anharmonicity of the qubit. 
To quantitatively understand the enhancement in the transduction 
and the asymmetry in spectral density with respect to $\tilde{\Delta}_q$,
we compute the cavity spectrum from Eq.~\ref{spectrum equation} 
as a function of susceptibilities. 
Approximately, the cavity spectral density can be written as,
\begin{align}
S_{a}(\omega) &\approx n_m^i \gamma_m (|\chi_m|^2 + |\tilde{\chi}_m|^2) \sigma(\omega), \text{where}\\
\sigma(\omega) &= \bigg|\frac{g J \chi_c \chi_{q\tilde{q}}(\Delta_q - 2 \eta)}{\Delta_q + 2 {i} \eta^2 \chi_{q\tilde{q}} + g^2 \chi_{m\tilde{m}} \chi_{q\tilde{q}}(\Delta_q - 2 \eta)}\bigg|^2 \\
\chi_{q\tilde{q}} &= \chi_q - \tilde{\chi}_q \\
\chi_{m\tilde{m}} &= \chi_m - \tilde{\chi}_m.
\end{align}
Here, we note that the presence of the effective anharmonicity $\eta$ 
in the above equation accounts for the asymmetry observed with 
respect to the detuning of qubit drive.
In the limit $\eta\to 0$, the expression of $\sigma$ becomes symmetric
with respect to $\Delta_q$ as it enters the expression through $\chi_{q\tilde{q}}$ only.

Similar to the measurement performed, we analyze the cavity 
spectral density as $\tilde{\Delta}_q$ is varied. 
Fig.~\ref{fig6}(a) shows theoretically calculated $\tilde{S}_a(\omega)$
using the device parameters.
We observe a pattern in $S_a(\omega)$ which is similar to the experimental
measurement.
For a quantitative comparison, we define the integrated spectrum 
as $S_T =\int S_{a}(\omega) d\omega$ 
and  evaluate it for experimental data.
Fig.~\ref{fig6}(b) shows the plot of $S_T$ from the experimental results
shown in Fig.~\ref{fig5}(c) and theoretical calculations. 
A good match validates the approximation made in arriving at 
the effective Hamiltonian in the theoretical calculations. 

%%%%%%%%%%% COMBED Till here

\section{Outlook and conclusion}\label{conclusion}

To summarize, this work has investigated a coupled three-mode 
hybrid system with a transmon qubit in the presence of external 
drives. Using the quantum noise and the Lindblad formalism, we study
the possibility of sideband cooling of the mechanical resonator by 
the qubit mode. We find that the readout of the mechanical 
mode is possible by coupling the transmon qubit to a readout
cavity just like in standard c-QED setup while maintaining 
a dispersive coupling between the cavity and the qubit.
In addition, we experimentally demonstrate the applicability of 
the readout scheme, wherein the experimental results
matches closely to the analytical calculations.
In this particular experiment, we do not observe 
any cooling of the mechanical resonator due to lower $g_0$ and low
sideband parameter $(\omega_m/\Gamma\approx~0.4)$. 
While the achieved flux responsivity of the qubit in
dispersive limit was high 16~GHz$/\Phi_0$, the estimated 
coupling 
rate ($g_0/2\pi\approx7.5$~kHz) was inadequate due to the lower applied 
magnetic field 1.1~mT.

\par
Looking ahead, the recent experiments have shown promising 
results for the transmon linewidth in the parallel magnetic field
up to hundreds of mT with no significant change in the spectroscopic linewidth \cite{krause_magnetic_2022}.
In addition, the flux responsivity of the qubit can be pushed to 
40~GHz$/\Phi_0$ by increasing the maximum qubit frequency. 
With these parameters, the single-photon electromechanical
coupling between qubit and mechanical resonator can 
be enhanced up to 10~MHz, bringing the 
system near to ultra-strong coupling 
regime \cite{forn-diaz_ultrastrong_2019}.
Such regime opens up the possibilities of observing the photon 
blockade effects \cite{rabl_photon_2011}, non-trivial ground 
state \cite{peterson_ultrastrong_2019}
and a path of using low frequency mechanical resonator in the 
quantum technologies.

\section{Acknowledgment}
The authors thank G.~S.~Agarwal and Manas~Kulkarni 
for valuable discussions. This material is based upon work 
supported by the Air Force Office of Scientific Research under 
award number FA2386-20-1-4003. 
V.S. acknowledge the support received under the Core Research 
Grant by the Department of Science and Technology (India). 
The authors acknowledge device fabrication facilities at CeNSE, IISc
Bangalore, and central facilities at the Department of
Physics funded by DST.

%\bibliography{ref}

\begin{thebibliography}{43}%
	\makeatletter
	\providecommand \@ifxundefined [1]{%
		\@ifx{#1\undefined}
	}%
	\providecommand \@ifnum [1]{%
		\ifnum #1\expandafter \@firstoftwo
		\else \expandafter \@secondoftwo
		\fi
	}%
	\providecommand \@ifx [1]{%
		\ifx #1\expandafter \@firstoftwo
		\else \expandafter \@secondoftwo
		\fi
	}%
	\providecommand \natexlab [1]{#1}%
	\providecommand \enquote  [1]{``#1''}%
	\providecommand \bibnamefont  [1]{#1}%
	\providecommand \bibfnamefont [1]{#1}%
	\providecommand \citenamefont [1]{#1}%
	\providecommand \href@noop [0]{\@secondoftwo}%
	\providecommand \href [0]{\begingroup \@sanitize@url \@href}%
	\providecommand \@href[1]{\@@startlink{#1}\@@href}%
	\providecommand \@@href[1]{\endgroup#1\@@endlink}%
	\providecommand \@sanitize@url [0]{\catcode `\\12\catcode `\$12\catcode
		`\&12\catcode `\#12\catcode `\^12\catcode `\_12\catcode `\%12\relax}%
	\providecommand \@@startlink[1]{}%
	\providecommand \@@endlink[0]{}%
	\providecommand \url  [0]{\begingroup\@sanitize@url \@url }%
	\providecommand \@url [1]{\endgroup\@href {#1}{\urlprefix }}%
	\providecommand \urlprefix  [0]{URL }%
	\providecommand \Eprint [0]{\href }%
	\providecommand \doibase [0]{https://doi.org/}%
	\providecommand \selectlanguage [0]{\@gobble}%
	\providecommand \bibinfo  [0]{\@secondoftwo}%
	\providecommand \bibfield  [0]{\@secondoftwo}%
	\providecommand \translation [1]{[#1]}%
	\providecommand \BibitemOpen [0]{}%
	\providecommand \bibitemStop [0]{}%
	\providecommand \bibitemNoStop [0]{.\EOS\space}%
	\providecommand \EOS [0]{\spacefactor3000\relax}%
	\providecommand \BibitemShut  [1]{\csname bibitem#1\endcsname}%
	\let\auto@bib@innerbib\@empty
	%</preamble>
	\bibitem [{\citenamefont {Barzanjeh}\ \emph {et~al.}(2022)\citenamefont
		{Barzanjeh}, \citenamefont {Xuereb}, \citenamefont {Gröblacher},
		\citenamefont {Paternostro}, \citenamefont {Regal},\ and\ \citenamefont
		{Weig}}]{barzanjeh_optomechanics_2022}%
	\BibitemOpen
	\bibfield  {author} {\bibinfo {author} {\bibfnamefont {S.}~\bibnamefont
			{Barzanjeh}}, \bibinfo {author} {\bibfnamefont {A.}~\bibnamefont {Xuereb}},
		\bibinfo {author} {\bibfnamefont {S.}~\bibnamefont {Gröblacher}}, \bibinfo
		{author} {\bibfnamefont {M.}~\bibnamefont {Paternostro}}, \bibinfo {author}
		{\bibfnamefont {C.~A.}\ \bibnamefont {Regal}},\ and\ \bibinfo {author}
		{\bibfnamefont {E.~M.}\ \bibnamefont {Weig}},\ }\bibfield  {title} {\bibinfo
		{title} {Optomechanics for quantum technologies},\ }\href
	{https://doi.org/10.1038/s41567-021-01402-0} {\bibfield  {journal} {\bibinfo
			{journal} {Nature Physics}\ }\textbf {\bibinfo {volume} {18}},\ \bibinfo
		{pages} {15} (\bibinfo {year} {2022})}\BibitemShut {NoStop}%
	\bibitem [{\citenamefont {Aspelmeyer}\ \emph {et~al.}(2014)\citenamefont
		{Aspelmeyer}, \citenamefont {Kippenberg},\ and\ \citenamefont
		{Marquardt}}]{aspelmeyer_cavity_2014}%
	\BibitemOpen
	\bibfield  {author} {\bibinfo {author} {\bibfnamefont {M.}~\bibnamefont
			{Aspelmeyer}}, \bibinfo {author} {\bibfnamefont {T.~J.}\ \bibnamefont
			{Kippenberg}},\ and\ \bibinfo {author} {\bibfnamefont {F.}~\bibnamefont
			{Marquardt}},\ }\bibfield  {title} {\bibinfo {title} {Cavity optomechanics},\
	}\href {https://doi.org/10.1103/RevModPhys.86.1391} {\bibfield  {journal}
		{\bibinfo  {journal} {Reviews of Modern Physics}\ }\textbf {\bibinfo {volume}
			{86}},\ \bibinfo {pages} {1391} (\bibinfo {year} {2014})}\BibitemShut
	{NoStop}%
	\bibitem [{\citenamefont {Teufel}\ \emph {et~al.}(2011)\citenamefont {Teufel},
		\citenamefont {Donner}, \citenamefont {Li}, \citenamefont {Harlow},
		\citenamefont {Allman}, \citenamefont {Cicak}, \citenamefont {Sirois},
		\citenamefont {Whittaker}, \citenamefont {Lehnert},\ and\ \citenamefont
		{Simmonds}}]{teufel_sideband_2011}%
	\BibitemOpen
	\bibfield  {author} {\bibinfo {author} {\bibfnamefont {J.~D.}\ \bibnamefont
			{Teufel}}, \bibinfo {author} {\bibfnamefont {T.}~\bibnamefont {Donner}},
		\bibinfo {author} {\bibfnamefont {D.}~\bibnamefont {Li}}, \bibinfo {author}
		{\bibfnamefont {J.~W.}\ \bibnamefont {Harlow}}, \bibinfo {author}
		{\bibfnamefont {M.~S.}\ \bibnamefont {Allman}}, \bibinfo {author}
		{\bibfnamefont {K.}~\bibnamefont {Cicak}}, \bibinfo {author} {\bibfnamefont
			{A.~J.}\ \bibnamefont {Sirois}}, \bibinfo {author} {\bibfnamefont {J.~D.}\
			\bibnamefont {Whittaker}}, \bibinfo {author} {\bibfnamefont {K.~W.}\
			\bibnamefont {Lehnert}},\ and\ \bibinfo {author} {\bibfnamefont {R.~W.}\
			\bibnamefont {Simmonds}},\ }\bibfield  {title} {\bibinfo {title} {Sideband
			cooling of micromechanical motion to the quantum ground state},\ }\href
	{https://doi.org/10.1038/nature10261} {\bibfield  {journal} {\bibinfo
			{journal} {Nature}\ }\textbf {\bibinfo {volume} {475}},\ \bibinfo {pages}
		{359} (\bibinfo {year} {2011})}\BibitemShut {NoStop}%
	\bibitem [{\citenamefont {Chan}\ \emph {et~al.}(2011)\citenamefont {Chan},
		\citenamefont {Alegre}, \citenamefont {Safavi-Naeini}, \citenamefont {Hill},
		\citenamefont {Krause}, \citenamefont {Gröblacher}, \citenamefont
		{Aspelmeyer},\ and\ \citenamefont {Painter}}]{chan_laser_2011}%
	\BibitemOpen
	\bibfield  {author} {\bibinfo {author} {\bibfnamefont {J.}~\bibnamefont
			{Chan}}, \bibinfo {author} {\bibfnamefont {T.~P.~M.}\ \bibnamefont {Alegre}},
		\bibinfo {author} {\bibfnamefont {A.~H.}\ \bibnamefont {Safavi-Naeini}},
		\bibinfo {author} {\bibfnamefont {J.~T.}\ \bibnamefont {Hill}}, \bibinfo
		{author} {\bibfnamefont {A.}~\bibnamefont {Krause}}, \bibinfo {author}
		{\bibfnamefont {S.}~\bibnamefont {Gröblacher}}, \bibinfo {author}
		{\bibfnamefont {M.}~\bibnamefont {Aspelmeyer}},\ and\ \bibinfo {author}
		{\bibfnamefont {O.}~\bibnamefont {Painter}},\ }\bibfield  {title} {\bibinfo
		{title} {Laser cooling of a nanomechanical oscillator into its quantum ground
			state},\ }\href {https://doi.org/10.1038/nature10461} {\bibfield  {journal}
		{\bibinfo  {journal} {Nature}\ }\textbf {\bibinfo {volume} {478}},\ \bibinfo
		{pages} {89} (\bibinfo {year} {2011})}\BibitemShut {NoStop}%
	\bibitem [{\citenamefont {Wollman}\ \emph {et~al.}(2015)\citenamefont
		{Wollman}, \citenamefont {Lei}, \citenamefont {Weinstein}, \citenamefont
		{Suh}, \citenamefont {Kronwald}, \citenamefont {Marquardt}, \citenamefont
		{Clerk},\ and\ \citenamefont {Schwab}}]{wollman_quantum_2015}%
	\BibitemOpen
	\bibfield  {author} {\bibinfo {author} {\bibfnamefont {E.~E.}\ \bibnamefont
			{Wollman}}, \bibinfo {author} {\bibfnamefont {C.~U.}\ \bibnamefont {Lei}},
		\bibinfo {author} {\bibfnamefont {A.~J.}\ \bibnamefont {Weinstein}}, \bibinfo
		{author} {\bibfnamefont {J.}~\bibnamefont {Suh}}, \bibinfo {author}
		{\bibfnamefont {A.}~\bibnamefont {Kronwald}}, \bibinfo {author}
		{\bibfnamefont {F.}~\bibnamefont {Marquardt}}, \bibinfo {author}
		{\bibfnamefont {A.~A.}\ \bibnamefont {Clerk}},\ and\ \bibinfo {author}
		{\bibfnamefont {K.~C.}\ \bibnamefont {Schwab}},\ }\bibfield  {title}
	{\bibinfo {title} {Quantum squeezing of motion in a mechanical resonator},\
	}\href {https://doi.org/10.1126/science.aac5138} {\bibfield  {journal}
		{\bibinfo  {journal} {Science}\ }\textbf {\bibinfo {volume} {349}},\ \bibinfo
		{pages} {952} (\bibinfo {year} {2015})},\ \BibitemShut {NoStop}%
	\bibitem [{\citenamefont {Ockeloen-Korppi}\ \emph {et~al.}(2018)\citenamefont
		{Ockeloen-Korppi}, \citenamefont {Damskägg}, \citenamefont {Pirkkalainen},
		\citenamefont {Asjad}, \citenamefont {Clerk}, \citenamefont {Massel},
		\citenamefont {Woolley},\ and\ \citenamefont
		{Sillanpää}}]{ockeloen-korppi_stabilized_2018}%
	\BibitemOpen
	\bibfield  {author} {\bibinfo {author} {\bibfnamefont {C.~F.}\ \bibnamefont
			{Ockeloen-Korppi}}, \bibinfo {author} {\bibfnamefont {E.}~\bibnamefont
			{Damskägg}}, \bibinfo {author} {\bibfnamefont {J.-M.}\ \bibnamefont
			{Pirkkalainen}}, \bibinfo {author} {\bibfnamefont {M.}~\bibnamefont {Asjad}},
		\bibinfo {author} {\bibfnamefont {A.~A.}\ \bibnamefont {Clerk}}, \bibinfo
		{author} {\bibfnamefont {F.}~\bibnamefont {Massel}}, \bibinfo {author}
		{\bibfnamefont {M.~J.}\ \bibnamefont {Woolley}},\ and\ \bibinfo {author}
		{\bibfnamefont {M.~A.}\ \bibnamefont {Sillanpää}},\ }\bibfield  {title}
	{\bibinfo {title} {Stabilized entanglement of massive mechanical
			oscillators},\ }\href {https://doi.org/10.1038/s41586-018-0038-x} {\bibfield
		{journal} {\bibinfo  {journal} {Nature}\ }\textbf {\bibinfo {volume} {556}},\
		\bibinfo {pages} {478} (\bibinfo {year} {2018})}\BibitemShut {NoStop}%
	\bibitem [{\citenamefont {Peterson}\ \emph {et~al.}(2019)\citenamefont
		{Peterson}, \citenamefont {Kotler}, \citenamefont {Lecocq}, \citenamefont
		{Cicak}, \citenamefont {Jin}, \citenamefont {Simmonds}, \citenamefont
		{Aumentado},\ and\ \citenamefont {Teufel}}]{peterson_ultrastrong_2019}%
	\BibitemOpen
	\bibfield  {author} {\bibinfo {author} {\bibfnamefont {G.}~\bibnamefont
			{Peterson}}, \bibinfo {author} {\bibfnamefont {S.}~\bibnamefont {Kotler}},
		\bibinfo {author} {\bibfnamefont {F.}~\bibnamefont {Lecocq}}, \bibinfo
		{author} {\bibfnamefont {K.}~\bibnamefont {Cicak}}, \bibinfo {author}
		{\bibfnamefont {X.}~\bibnamefont {Jin}}, \bibinfo {author} {\bibfnamefont
			{R.}~\bibnamefont {Simmonds}}, \bibinfo {author} {\bibfnamefont
			{J.}~\bibnamefont {Aumentado}},\ and\ \bibinfo {author} {\bibfnamefont
			{J.}~\bibnamefont {Teufel}},\ }\bibfield  {title} {\bibinfo {title}
		{Ultrastrong {Parametric} {Coupling} between a {Superconducting} {Cavity} and
			a {Mechanical} {Resonator}},\ }\href
	{https://doi.org/10.1103/PhysRevLett.123.247701} {\bibfield  {journal}
		{\bibinfo  {journal} {Physical Review Letters}\ }\textbf {\bibinfo {volume}
			{123}},\ \bibinfo {pages} {247701} (\bibinfo {year} {2019})}\BibitemShut
	{NoStop}%
	\bibitem [{\citenamefont {Kotler}\ \emph {et~al.}(2021)\citenamefont {Kotler},
		\citenamefont {Peterson}, \citenamefont {Shojaee}, \citenamefont {Lecocq},
		\citenamefont {Cicak}, \citenamefont {Kwiatkowski}, \citenamefont {Geller},
		\citenamefont {Glancy}, \citenamefont {Knill}, \citenamefont {Simmonds},
		\citenamefont {Aumentado},\ and\ \citenamefont
		{Teufel}}]{kotler_direct_2021}%
	\BibitemOpen
	\bibfield  {author} {\bibinfo {author} {\bibfnamefont {S.}~\bibnamefont
			{Kotler}}, \bibinfo {author} {\bibfnamefont {G.~A.}\ \bibnamefont
			{Peterson}}, \bibinfo {author} {\bibfnamefont {E.}~\bibnamefont {Shojaee}},
		\bibinfo {author} {\bibfnamefont {F.}~\bibnamefont {Lecocq}}, \bibinfo
		{author} {\bibfnamefont {K.}~\bibnamefont {Cicak}}, \bibinfo {author}
		{\bibfnamefont {A.}~\bibnamefont {Kwiatkowski}}, \bibinfo {author}
		{\bibfnamefont {S.}~\bibnamefont {Geller}}, \bibinfo {author} {\bibfnamefont
			{S.}~\bibnamefont {Glancy}}, \bibinfo {author} {\bibfnamefont
			{E.}~\bibnamefont {Knill}}, \bibinfo {author} {\bibfnamefont {R.~W.}\
			\bibnamefont {Simmonds}}, \bibinfo {author} {\bibfnamefont {J.}~\bibnamefont
			{Aumentado}},\ and\ \bibinfo {author} {\bibfnamefont {J.~D.}\ \bibnamefont
			{Teufel}},\ }\bibfield  {title} {\bibinfo {title} {Direct observation of
			deterministic macroscopic entanglement},\ }\href
	{https://doi.org/10.1126/science.abf2998} {\bibfield  {journal} {\bibinfo
			{journal} {Science}\ }\textbf {\bibinfo {volume} {372}},\ \bibinfo {pages}
		{622} (\bibinfo {year} {2021})}\BibitemShut {NoStop}%
	\bibitem [{\citenamefont {Wollack}\ \emph {et~al.}(2022)\citenamefont
		{Wollack}, \citenamefont {Cleland}, \citenamefont {Gruenke}, \citenamefont
		{Wang}, \citenamefont {Arrangoiz-Arriola},\ and\ \citenamefont
		{Safavi-Naeini}}]{wollack_quantum_2022}%
	\BibitemOpen
	\bibfield  {author} {\bibinfo {author} {\bibfnamefont {E.~A.}\ \bibnamefont
			{Wollack}}, \bibinfo {author} {\bibfnamefont {A.~Y.}\ \bibnamefont
			{Cleland}}, \bibinfo {author} {\bibfnamefont {R.~G.}\ \bibnamefont
			{Gruenke}}, \bibinfo {author} {\bibfnamefont {Z.}~\bibnamefont {Wang}},
		\bibinfo {author} {\bibfnamefont {P.}~\bibnamefont {Arrangoiz-Arriola}},\
		and\ \bibinfo {author} {\bibfnamefont {A.~H.}\ \bibnamefont
			{Safavi-Naeini}},\ }\bibfield  {title} {\bibinfo {title} {Quantum state
			preparation and tomography of entangled mechanical resonators},\ }\href
	{https://doi.org/10.1038/s41586-022-04500-y} {\bibfield  {journal} {\bibinfo
			{journal} {Nature}\ }\textbf {\bibinfo {volume} {604}},\ \bibinfo {pages}
		{463} (\bibinfo {year} {2022})}\BibitemShut {NoStop}%
	\bibitem [{\citenamefont {Rabl}(2011)}]{rabl_photon_2011}%
	\BibitemOpen
	\bibfield  {author} {\bibinfo {author} {\bibfnamefont {P.}~\bibnamefont
			{Rabl}},\ }\bibfield  {title} {\bibinfo {title} {Photon {Blockade} {Effect}
			in {Optomechanical} {Systems}},\ }\href
	{https://doi.org/10.1103/PhysRevLett.107.063601} {\bibfield  {journal}
		{\bibinfo  {journal} {Physical Review Letters}\ }\textbf {\bibinfo {volume}
			{107}},\ \bibinfo {pages} {063601} (\bibinfo {year} {2011})}\BibitemShut
	{NoStop}%
	\bibitem [{\citenamefont {Nunnenkamp}\ \emph {et~al.}(2011)\citenamefont
		{Nunnenkamp}, \citenamefont {Børkje},\ and\ \citenamefont
		{Girvin}}]{nunnenkamp_single-photon_2011}%
	\BibitemOpen
	\bibfield  {author} {\bibinfo {author} {\bibfnamefont {A.}~\bibnamefont
			{Nunnenkamp}}, \bibinfo {author} {\bibfnamefont {K.}~\bibnamefont
			{Børkje}},\ and\ \bibinfo {author} {\bibfnamefont {S.~M.}\ \bibnamefont
			{Girvin}},\ }\bibfield  {title} {\bibinfo {title} {Single-{Photon}
			{Optomechanics}},\ }\href {https://doi.org/10.1103/PhysRevLett.107.063602}
	{\bibfield  {journal} {\bibinfo  {journal} {Physical Review Letters}\
		}\textbf {\bibinfo {volume} {107}},\ \bibinfo {pages} {063602} (\bibinfo
		{year} {2011})}\BibitemShut {NoStop}%
	\bibitem [{\citenamefont {Rabl}(2010)}]{rabl_cooling_2010}%
	\BibitemOpen
	\bibfield  {author} {\bibinfo {author} {\bibfnamefont {P.}~\bibnamefont
			{Rabl}},\ }\bibfield  {title} {\bibinfo {title} {Cooling of mechanical motion
			with a two-level system: {The} high-temperature regime},\ }\href
	{https://doi.org/10.1103/PhysRevB.82.165320} {\bibfield  {journal} {\bibinfo
			{journal} {Physical Review B}\ }\textbf {\bibinfo {volume} {82}},\ \bibinfo
		{pages} {165320} (\bibinfo {year} {2010})}\BibitemShut {NoStop}%
	\bibitem [{\citenamefont {Xiang}\ \emph {et~al.}(2013)\citenamefont {Xiang},
		\citenamefont {Ashhab}, \citenamefont {You},\ and\ \citenamefont
		{Nori}}]{xiang_hybrid_2013}%
	\BibitemOpen
	\bibfield  {author} {\bibinfo {author} {\bibfnamefont {Z.-L.}\ \bibnamefont
			{Xiang}}, \bibinfo {author} {\bibfnamefont {S.}~\bibnamefont {Ashhab}},
		\bibinfo {author} {\bibfnamefont {J.~Q.}\ \bibnamefont {You}},\ and\ \bibinfo
		{author} {\bibfnamefont {F.}~\bibnamefont {Nori}},\ }\bibfield  {title}
	{\bibinfo {title} {Hybrid quantum circuits: {Superconducting} circuits
			interacting with other quantum systems},\ }\href
	{https://doi.org/10.1103/RevModPhys.85.623} {\bibfield  {journal} {\bibinfo
			{journal} {Reviews of Modern Physics}\ }\textbf {\bibinfo {volume} {85}},\
		\bibinfo {pages} {623} (\bibinfo {year} {2013})}\BibitemShut {NoStop}%
	\bibitem [{\citenamefont {Clerk}\ \emph {et~al.}(2020)\citenamefont {Clerk},
		\citenamefont {Lehnert}, \citenamefont {Bertet}, \citenamefont {Petta},\ and\
		\citenamefont {Nakamura}}]{clerk_hybrid_2020}%
	\BibitemOpen
	\bibfield  {author} {\bibinfo {author} {\bibfnamefont {A.~A.}\ \bibnamefont
			{Clerk}}, \bibinfo {author} {\bibfnamefont {K.~W.}\ \bibnamefont {Lehnert}},
		\bibinfo {author} {\bibfnamefont {P.}~\bibnamefont {Bertet}}, \bibinfo
		{author} {\bibfnamefont {J.~R.}\ \bibnamefont {Petta}},\ and\ \bibinfo
		{author} {\bibfnamefont {Y.}~\bibnamefont {Nakamura}},\ }\bibfield  {title}
	{\bibinfo {title} {Hybrid quantum systems with circuit quantum
			electrodynamics},\ }\href {https://doi.org/10.1038/s41567-020-0797-9}
	{\bibfield  {journal} {\bibinfo  {journal} {Nature Physics}\ }\textbf
		{\bibinfo {volume} {16}},\ \bibinfo {pages} {257} (\bibinfo {year}
		{2020})}\BibitemShut {NoStop}%
	\bibitem [{\citenamefont {Martin}\ \emph {et~al.}(2004)\citenamefont {Martin},
		\citenamefont {Shnirman}, \citenamefont {Tian},\ and\ \citenamefont
		{Zoller}}]{martin_ground-state_2004}%
	\BibitemOpen
	\bibfield  {author} {\bibinfo {author} {\bibfnamefont {I.}~\bibnamefont
			{Martin}}, \bibinfo {author} {\bibfnamefont {A.}~\bibnamefont {Shnirman}},
		\bibinfo {author} {\bibfnamefont {L.}~\bibnamefont {Tian}},\ and\ \bibinfo
		{author} {\bibfnamefont {P.}~\bibnamefont {Zoller}},\ }\bibfield  {title}
	{\bibinfo {title} {Ground-state cooling of mechanical resonators},\ }\href
	{https://doi.org/10.1103/PhysRevB.69.125339} {\bibfield  {journal} {\bibinfo
			{journal} {Physical Review B}\ }\textbf {\bibinfo {volume} {69}},\ \bibinfo
		{pages} {125339} (\bibinfo {year} {2004})}\BibitemShut {NoStop}%
	\bibitem [{\citenamefont {Khosla}\ \emph {et~al.}(2018)\citenamefont {Khosla},
		\citenamefont {Vanner}, \citenamefont {Ares},\ and\ \citenamefont
		{Laird}}]{khosla_displacemon_2018}%
	\BibitemOpen
	\bibfield  {author} {\bibinfo {author} {\bibfnamefont {K.}~\bibnamefont
			{Khosla}}, \bibinfo {author} {\bibfnamefont {M.}~\bibnamefont {Vanner}},
		\bibinfo {author} {\bibfnamefont {N.}~\bibnamefont {Ares}},\ and\ \bibinfo
		{author} {\bibfnamefont {E.}~\bibnamefont {Laird}},\ }\bibfield  {title}
	{\bibinfo {title} {Displacemon {Electromechanics}: {How} to {Detect}
			{Quantum} {Interference} in a {Nanomechanical} {Resonator}},\ }\href
	{https://doi.org/10.1103/PhysRevX.8.021052} {\bibfield  {journal} {\bibinfo
			{journal} {Physical Review X}\ }\textbf {\bibinfo {volume} {8}},\ \bibinfo
		{pages} {021052} (\bibinfo {year} {2018})}\BibitemShut {NoStop}%
	\bibitem [{\citenamefont {Jaehne}\ \emph {et~al.}(2008)\citenamefont {Jaehne},
		\citenamefont {Hammerer},\ and\ \citenamefont
		{Wallquist}}]{jaehne_ground-state_2008}%
	\BibitemOpen
	\bibfield  {author} {\bibinfo {author} {\bibfnamefont {K.}~\bibnamefont
			{Jaehne}}, \bibinfo {author} {\bibfnamefont {K.}~\bibnamefont {Hammerer}},\
		and\ \bibinfo {author} {\bibfnamefont {M.}~\bibnamefont {Wallquist}},\
	}\bibfield  {title} {\bibinfo {title} {Ground-state cooling of a
			nanomechanical resonator via a {Cooper}-pair box qubit},\ }\href
	{https://doi.org/10.1088/1367-2630/10/9/095019} {\bibfield  {journal}
		{\bibinfo  {journal} {New Journal of Physics}\ }\textbf {\bibinfo {volume}
			{10}},\ \bibinfo {pages} {095019} (\bibinfo {year} {2008})}\BibitemShut
	{NoStop}%
	\bibitem [{\citenamefont {Hauss}\ \emph {et~al.}(2008)\citenamefont {Hauss},
		\citenamefont {Fedorov}, \citenamefont {André}, \citenamefont {Brosco},
		\citenamefont {Hutter}, \citenamefont {Kothari}, \citenamefont {Yeshwanth},
		\citenamefont {Shnirman},\ and\ \citenamefont
		{Schön}}]{hauss_dissipation_2008}%
	\BibitemOpen
	\bibfield  {author} {\bibinfo {author} {\bibfnamefont {J.}~\bibnamefont
			{Hauss}}, \bibinfo {author} {\bibfnamefont {A.}~\bibnamefont {Fedorov}},
		\bibinfo {author} {\bibfnamefont {S.}~\bibnamefont {André}}, \bibinfo
		{author} {\bibfnamefont {V.}~\bibnamefont {Brosco}}, \bibinfo {author}
		{\bibfnamefont {C.}~\bibnamefont {Hutter}}, \bibinfo {author} {\bibfnamefont
			{R.}~\bibnamefont {Kothari}}, \bibinfo {author} {\bibfnamefont
			{S.}~\bibnamefont {Yeshwanth}}, \bibinfo {author} {\bibfnamefont
			{A.}~\bibnamefont {Shnirman}},\ and\ \bibinfo {author} {\bibfnamefont
			{G.}~\bibnamefont {Schön}},\ }\bibfield  {title} {\bibinfo {title}
		{Dissipation in circuit quantum electrodynamics: lasing and cooling of a
			low-frequency oscillator},\ }\href
	{https://doi.org/10.1088/1367-2630/10/9/095018} {\bibfield  {journal}
		{\bibinfo  {journal} {New Journal of Physics}\ }\textbf {\bibinfo {volume}
			{10}},\ \bibinfo {pages} {095018} (\bibinfo {year} {2008})}\BibitemShut
	{NoStop}%
	\bibitem [{\citenamefont {Wang}\ \emph {et~al.}(2009)\citenamefont {Wang},
		\citenamefont {Li}, \citenamefont {Xue}, \citenamefont {Bruder},\ and\
		\citenamefont {Semba}}]{wang_cooling_2009}%
	\BibitemOpen
	\bibfield  {author} {\bibinfo {author} {\bibfnamefont {Y.-D.}\ \bibnamefont
			{Wang}}, \bibinfo {author} {\bibfnamefont {Y.}~\bibnamefont {Li}}, \bibinfo
		{author} {\bibfnamefont {F.}~\bibnamefont {Xue}}, \bibinfo {author}
		{\bibfnamefont {C.}~\bibnamefont {Bruder}},\ and\ \bibinfo {author}
		{\bibfnamefont {K.}~\bibnamefont {Semba}},\ }\bibfield  {title} {\bibinfo
		{title} {Cooling a micromechanical resonator by quantum back-action from a
			noisy qubit},\ }\href {https://doi.org/10.1103/PhysRevB.80.144508} {\bibfield
		{journal} {\bibinfo  {journal} {Physical Review B}\ }\textbf {\bibinfo
			{volume} {80}},\ \bibinfo {pages} {144508} (\bibinfo {year}
		{2009})}\BibitemShut {NoStop}%
	\bibitem [{\citenamefont {Nongthombam}\ \emph {et~al.}(2021)\citenamefont
		{Nongthombam}, \citenamefont {Sahoo},\ and\ \citenamefont
		{Sarma}}]{nongthombam_ground-state_2021}%
	\BibitemOpen
	\bibfield  {author} {\bibinfo {author} {\bibfnamefont {R.}~\bibnamefont
			{Nongthombam}}, \bibinfo {author} {\bibfnamefont {A.}~\bibnamefont {Sahoo}},\
		and\ \bibinfo {author} {\bibfnamefont {A.~K.}\ \bibnamefont {Sarma}},\
	}\bibfield  {title} {\bibinfo {title} {Ground-state cooling of a mechanical
			oscillator via a hybrid electro-optomechanical system},\ }\href
	{https://doi.org/10.1103/PhysRevA.104.023509} {\bibfield  {journal} {\bibinfo
			{journal} {Physical Review A}\ }\textbf {\bibinfo {volume} {104}},\ \bibinfo
		{pages} {023509} (\bibinfo {year} {2021})}\BibitemShut {NoStop}%
	\bibitem [{\citenamefont {Wang}\ \emph {et~al.}(2018)\citenamefont {Wang},
		\citenamefont {Miranowicz}, \citenamefont {Li}, \citenamefont {Li},\ and\
		\citenamefont {Nori}}]{wang_two-color_2018}%
	\BibitemOpen
	\bibfield  {author} {\bibinfo {author} {\bibfnamefont {X.}~\bibnamefont
			{Wang}}, \bibinfo {author} {\bibfnamefont {A.}~\bibnamefont {Miranowicz}},
		\bibinfo {author} {\bibfnamefont {H.-R.}\ \bibnamefont {Li}}, \bibinfo
		{author} {\bibfnamefont {F.-L.}\ \bibnamefont {Li}},\ and\ \bibinfo {author}
		{\bibfnamefont {F.}~\bibnamefont {Nori}},\ }\bibfield  {title} {\bibinfo
		{title} {Two-color electromagnetically induced transparency via modulated
			coupling between a mechanical resonator and a qubit},\ }\href
	{https://doi.org/10.1103/PhysRevA.98.023821} {\bibfield  {journal} {\bibinfo
			{journal} {Physical Review A}\ }\textbf {\bibinfo {volume} {98}},\ \bibinfo
		{pages} {023821} (\bibinfo {year} {2018})}\BibitemShut {NoStop}%
	\bibitem [{\citenamefont {Manninen}\ \emph {et~al.}(2022)\citenamefont
		{Manninen}, \citenamefont {Haque}, \citenamefont {Vitali},\ and\
		\citenamefont {Hakonen}}]{manninen_enhancement_2022}%
	\BibitemOpen
	\bibfield  {author} {\bibinfo {author} {\bibfnamefont {J.}~\bibnamefont
			{Manninen}}, \bibinfo {author} {\bibfnamefont {M.~T.}\ \bibnamefont {Haque}},
		\bibinfo {author} {\bibfnamefont {D.}~\bibnamefont {Vitali}},\ and\ \bibinfo
		{author} {\bibfnamefont {P.}~\bibnamefont {Hakonen}},\ }\bibfield  {title}
	{\bibinfo {title} {Enhancement of the optomechanical coupling and {Kerr}
			nonlinearity using the {Josephson} capacitance of a {Cooper}-pair box},\
	}\href {https://doi.org/10.1103/PhysRevB.105.144508} {\bibfield  {journal}
		{\bibinfo  {journal} {Physical Review B}\ }\textbf {\bibinfo {volume}
			{105}},\ \bibinfo {pages} {144508} (\bibinfo {year} {2022})}\BibitemShut
	{NoStop}%
	\bibitem [{\citenamefont {Wilson-Rae}\ \emph {et~al.}(2004)\citenamefont
		{Wilson-Rae}, \citenamefont {Zoller},\ and\ \citenamefont
		{Imamoḡlu}}]{wilson-rae_laser_2004}%
	\BibitemOpen
	\bibfield  {author} {\bibinfo {author} {\bibfnamefont {I.}~\bibnamefont
			{Wilson-Rae}}, \bibinfo {author} {\bibfnamefont {P.}~\bibnamefont {Zoller}},\
		and\ \bibinfo {author} {\bibfnamefont {A.}~\bibnamefont {Imamoḡlu}},\
	}\bibfield  {title} {\bibinfo {title} {Laser {Cooling} of a {Nanomechanical}
			{Resonator} {Mode} to its {Quantum} {Ground} {State}},\ }\href
	{https://link.aps.org/doi/10.1103/PhysRevLett.92.075507} {\bibfield
		{journal} {\bibinfo  {journal} {Physical Review Letters}\ }\textbf {\bibinfo
			{volume} {92}},\ \bibinfo {pages} {075507} (\bibinfo {year}
		{2004})}\BibitemShut {NoStop}%
	\bibitem [{\citenamefont {Rabl}\ \emph {et~al.}(2009)\citenamefont {Rabl},
		\citenamefont {Cappellaro}, \citenamefont {Dutt}, \citenamefont {Jiang},
		\citenamefont {Maze},\ and\ \citenamefont {Lukin}}]{rabl_strong_2009}%
	\BibitemOpen
	\bibfield  {author} {\bibinfo {author} {\bibfnamefont {P.}~\bibnamefont
			{Rabl}}, \bibinfo {author} {\bibfnamefont {P.}~\bibnamefont {Cappellaro}},
		\bibinfo {author} {\bibfnamefont {M.~V.~G.}\ \bibnamefont {Dutt}}, \bibinfo
		{author} {\bibfnamefont {L.}~\bibnamefont {Jiang}}, \bibinfo {author}
		{\bibfnamefont {J.~R.}\ \bibnamefont {Maze}},\ and\ \bibinfo {author}
		{\bibfnamefont {M.~D.}\ \bibnamefont {Lukin}},\ }\bibfield  {title} {\bibinfo
		{title} {Strong magnetic coupling between an electronic spin qubit and a
			mechanical resonator},\ }\href {https://doi.org/10.1103/PhysRevB.79.041302}
	{\bibfield  {journal} {\bibinfo  {journal} {Physical Review B}\ }\textbf
		{\bibinfo {volume} {79}},\ \bibinfo {pages} {041302} (\bibinfo {year}
		{2009})}\BibitemShut {NoStop}%
	\bibitem [{\citenamefont {Pirkkalainen}\ \emph {et~al.}(2015)\citenamefont
		{Pirkkalainen}, \citenamefont {Cho}, \citenamefont {Massel}, \citenamefont
		{Tuorila}, \citenamefont {Heikkilä}, \citenamefont {Hakonen},\ and\
		\citenamefont {Sillanpää}}]{pirkkalainen_cavity_2015}%
	\BibitemOpen
	\bibfield  {author} {\bibinfo {author} {\bibfnamefont {J.-M.}\ \bibnamefont
			{Pirkkalainen}}, \bibinfo {author} {\bibfnamefont {S.~U.}\ \bibnamefont
			{Cho}}, \bibinfo {author} {\bibfnamefont {F.}~\bibnamefont {Massel}},
		\bibinfo {author} {\bibfnamefont {J.}~\bibnamefont {Tuorila}}, \bibinfo
		{author} {\bibfnamefont {T.~T.}\ \bibnamefont {Heikkilä}}, \bibinfo {author}
		{\bibfnamefont {P.~J.}\ \bibnamefont {Hakonen}},\ and\ \bibinfo {author}
		{\bibfnamefont {M.~A.}\ \bibnamefont {Sillanpää}},\ }\bibfield  {title}
	{\bibinfo {title} {Cavity optomechanics mediated by a quantum two-level
			system},\ }\href {https://doi.org/10.1038/ncomms7981} {\bibfield  {journal}
		{\bibinfo  {journal} {Nature Communications}\ }\textbf {\bibinfo {volume}
			{6}},\ \bibinfo {pages} {6981} (\bibinfo {year} {2015})}\BibitemShut
	{NoStop}%
	\bibitem [{\citenamefont {Pirkkalainen}\ \emph {et~al.}(2013)\citenamefont
		{Pirkkalainen}, \citenamefont {Cho}, \citenamefont {Li}, \citenamefont
		{Paraoanu}, \citenamefont {Hakonen},\ and\ \citenamefont
		{Sillanpää}}]{pirkkalainen_hybrid_2013}%
	\BibitemOpen
	\bibfield  {author} {\bibinfo {author} {\bibfnamefont {J.-M.}\ \bibnamefont
			{Pirkkalainen}}, \bibinfo {author} {\bibfnamefont {S.~U.}\ \bibnamefont
			{Cho}}, \bibinfo {author} {\bibfnamefont {J.}~\bibnamefont {Li}}, \bibinfo
		{author} {\bibfnamefont {G.~S.}\ \bibnamefont {Paraoanu}}, \bibinfo {author}
		{\bibfnamefont {P.~J.}\ \bibnamefont {Hakonen}},\ and\ \bibinfo {author}
		{\bibfnamefont {M.~A.}\ \bibnamefont {Sillanpää}},\ }\bibfield  {title}
	{\bibinfo {title} {Hybrid circuit cavity quantum electrodynamics with a
			micromechanical resonator},\ }\href {https://doi.org/10.1038/nature11821}
	{\bibfield  {journal} {\bibinfo  {journal} {Nature}\ }\textbf {\bibinfo
			{volume} {494}},\ \bibinfo {pages} {211} (\bibinfo {year}
		{2013})}\BibitemShut {NoStop}%
	\bibitem [{\citenamefont {Viennot}\ \emph {et~al.}(2018)\citenamefont
		{Viennot}, \citenamefont {Ma},\ and\ \citenamefont
		{Lehnert}}]{viennot_phonon-number-sensitive_2018}%
	\BibitemOpen
	\bibfield  {author} {\bibinfo {author} {\bibfnamefont {J.}~\bibnamefont
			{Viennot}}, \bibinfo {author} {\bibfnamefont {X.}~\bibnamefont {Ma}},\ and\
		\bibinfo {author} {\bibfnamefont {K.}~\bibnamefont {Lehnert}},\ }\bibfield
	{title} {\bibinfo {title} {Phonon-{Number}-{Sensitive} {Electromechanics}},\
	}\href {https://doi.org/10.1103/PhysRevLett.121.183601} {\bibfield  {journal}
		{\bibinfo  {journal} {Physical Review Letters}\ }\textbf {\bibinfo {volume}
			{121}},\ \bibinfo {pages} {183601} (\bibinfo {year} {2018})}\BibitemShut
	{NoStop}%
	\bibitem [{\citenamefont {Rodrigues}\ \emph {et~al.}(2019)\citenamefont
		{Rodrigues}, \citenamefont {Bothner},\ and\ \citenamefont
		{Steele}}]{rodrigues_coupling_2019}%
	\BibitemOpen
	\bibfield  {author} {\bibinfo {author} {\bibfnamefont {I.~C.}\ \bibnamefont
			{Rodrigues}}, \bibinfo {author} {\bibfnamefont {D.}~\bibnamefont {Bothner}},\
		and\ \bibinfo {author} {\bibfnamefont {G.~A.}\ \bibnamefont {Steele}},\
	}\bibfield  {title} {\bibinfo {title} {Coupling microwave photons to a
			mechanical resonator using quantum interference},\ }\bibfield  {journal}
	{\bibinfo  {journal} {Nature Communications}\ }\textbf {\bibinfo {volume}
		{10}},\ \href {https://doi.org/10.1038/s41467-019-12964-2}
	{10.1038/s41467-019-12964-2} (\bibinfo {year} {2019})\BibitemShut {NoStop}%
	\bibitem [{\citenamefont {Schmidt}\ \emph {et~al.}(2020)\citenamefont
		{Schmidt}, \citenamefont {T.~Amawi}, \citenamefont {Pogorzalek},
		\citenamefont {Deppe}, \citenamefont {Marx}, \citenamefont {Gross},\ and\
		\citenamefont {Huebl}}]{schmidt_sideband-resolved_2020}%
	\BibitemOpen
	\bibfield  {author} {\bibinfo {author} {\bibfnamefont {P.}~\bibnamefont
			{Schmidt}}, \bibinfo {author} {\bibfnamefont {M.}~\bibnamefont {T.~Amawi}},
		\bibinfo {author} {\bibfnamefont {S.}~\bibnamefont {Pogorzalek}}, \bibinfo
		{author} {\bibfnamefont {F.}~\bibnamefont {Deppe}}, \bibinfo {author}
		{\bibfnamefont {A.}~\bibnamefont {Marx}}, \bibinfo {author} {\bibfnamefont
			{R.}~\bibnamefont {Gross}},\ and\ \bibinfo {author} {\bibfnamefont
			{H.}~\bibnamefont {Huebl}},\ }\bibfield  {title} {\bibinfo {title}
		{Sideband-resolved resonator electromechanics based on a nonlinear
			{Josephson} inductance probed on the single-photon level},\ }\href
	{https://doi.org/10.1038/s42005-020-00501-3} {\bibfield  {journal} {\bibinfo
			{journal} {Communications Physics}\ }\textbf {\bibinfo {volume} {3}},\
		\bibinfo {pages} {1} (\bibinfo {year} {2020})}\BibitemShut {NoStop}%
	\bibitem [{\citenamefont {Zoepfl}\ \emph {et~al.}(2020)\citenamefont {Zoepfl},
		\citenamefont {Juan}, \citenamefont {Schneider},\ and\ \citenamefont
		{Kirchmair}}]{zoepfl_single-photon_2020}%
	\BibitemOpen
	\bibfield  {author} {\bibinfo {author} {\bibfnamefont {D.}~\bibnamefont
			{Zoepfl}}, \bibinfo {author} {\bibfnamefont {M.}~\bibnamefont {Juan}},
		\bibinfo {author} {\bibfnamefont {C.}~\bibnamefont {Schneider}},\ and\
		\bibinfo {author} {\bibfnamefont {G.}~\bibnamefont {Kirchmair}},\ }\bibfield
	{title} {\bibinfo {title} {Single-{Photon} {Cooling} in {Microwave}
			{Magnetomechanics}},\ }\href {https://doi.org/10.1103/PhysRevLett.125.023601}
	{\bibfield  {journal} {\bibinfo  {journal} {Physical Review Letters}\
		}\textbf {\bibinfo {volume} {125}},\ \bibinfo {pages} {023601} (\bibinfo
		{year} {2020})}\BibitemShut {NoStop}%
	\bibitem [{\citenamefont {Bera}\ \emph {et~al.}(2021)\citenamefont {Bera},
		\citenamefont {Majumder}, \citenamefont {Sahu},\ and\ \citenamefont
		{Singh}}]{bera_large_2021}%
	\BibitemOpen
	\bibfield  {author} {\bibinfo {author} {\bibfnamefont {T.}~\bibnamefont
			{Bera}}, \bibinfo {author} {\bibfnamefont {S.}~\bibnamefont {Majumder}},
		\bibinfo {author} {\bibfnamefont {S.~K.}\ \bibnamefont {Sahu}},\ and\
		\bibinfo {author} {\bibfnamefont {V.}~\bibnamefont {Singh}},\ }\bibfield
	{title} {\bibinfo {title} {Large flux-mediated coupling in hybrid
			electromechanical system with a transmon qubit},\ }\bibfield  {journal}
	{\bibinfo  {journal} {Communications Physics}\ }\textbf {\bibinfo {volume}
		{4}},\ \href {https://doi.org/10.1038/s42005-020-00514-y}
	{10.1038/s42005-020-00514-y} (\bibinfo {year} {2021})\BibitemShut {NoStop}%
	\bibitem [{\citenamefont {O’Connell}\ \emph {et~al.}(2010)\citenamefont
		{O’Connell}, \citenamefont {Hofheinz}, \citenamefont {Ansmann},
		\citenamefont {Bialczak}, \citenamefont {Lenander}, \citenamefont {Lucero},
		\citenamefont {Neeley}, \citenamefont {Sank}, \citenamefont {Wang},
		\citenamefont {Weides}, \citenamefont {Wenner}, \citenamefont {Martinis},\
		and\ \citenamefont {Cleland}}]{oconnell_quantum_2010}%
	\BibitemOpen
	\bibfield  {author} {\bibinfo {author} {\bibfnamefont {A.~D.}\ \bibnamefont
			{O’Connell}}, \bibinfo {author} {\bibfnamefont {M.}~\bibnamefont
			{Hofheinz}}, \bibinfo {author} {\bibfnamefont {M.}~\bibnamefont {Ansmann}},
		\bibinfo {author} {\bibfnamefont {R.~C.}\ \bibnamefont {Bialczak}}, \bibinfo
		{author} {\bibfnamefont {M.}~\bibnamefont {Lenander}}, \bibinfo {author}
		{\bibfnamefont {E.}~\bibnamefont {Lucero}}, \bibinfo {author} {\bibfnamefont
			{M.}~\bibnamefont {Neeley}}, \bibinfo {author} {\bibfnamefont
			{D.}~\bibnamefont {Sank}}, \bibinfo {author} {\bibfnamefont {H.}~\bibnamefont
			{Wang}}, \bibinfo {author} {\bibfnamefont {M.}~\bibnamefont {Weides}},
		\bibinfo {author} {\bibfnamefont {J.}~\bibnamefont {Wenner}}, \bibinfo
		{author} {\bibfnamefont {J.~M.}\ \bibnamefont {Martinis}},\ and\ \bibinfo
		{author} {\bibfnamefont {A.~N.}\ \bibnamefont {Cleland}},\ }\bibfield
	{title} {\bibinfo {title} {Quantum ground state and single-phonon control of
			a mechanical resonator},\ }\href {https://doi.org/10.1038/nature08967}
	{\bibfield  {journal} {\bibinfo  {journal} {Nature}\ }\textbf {\bibinfo
			{volume} {464}},\ \bibinfo {pages} {697} (\bibinfo {year}
		{2010})}\BibitemShut {NoStop}%
	\bibitem [{\citenamefont {Arrangoiz-Arriola}\ \emph {et~al.}(2019)\citenamefont
		{Arrangoiz-Arriola}, \citenamefont {Wollack}, \citenamefont {Wang},
		\citenamefont {Pechal}, \citenamefont {Jiang}, \citenamefont {McKenna},
		\citenamefont {Witmer}, \citenamefont {Van~Laer},\ and\ \citenamefont
		{Safavi-Naeini}}]{arrangoiz-arriola_resolving_2019}%
	\BibitemOpen
	\bibfield  {author} {\bibinfo {author} {\bibfnamefont {P.}~\bibnamefont
			{Arrangoiz-Arriola}}, \bibinfo {author} {\bibfnamefont {E.~A.}\ \bibnamefont
			{Wollack}}, \bibinfo {author} {\bibfnamefont {Z.}~\bibnamefont {Wang}},
		\bibinfo {author} {\bibfnamefont {M.}~\bibnamefont {Pechal}}, \bibinfo
		{author} {\bibfnamefont {W.}~\bibnamefont {Jiang}}, \bibinfo {author}
		{\bibfnamefont {T.~P.}\ \bibnamefont {McKenna}}, \bibinfo {author}
		{\bibfnamefont {J.~D.}\ \bibnamefont {Witmer}}, \bibinfo {author}
		{\bibfnamefont {R.}~\bibnamefont {Van~Laer}},\ and\ \bibinfo {author}
		{\bibfnamefont {A.~H.}\ \bibnamefont {Safavi-Naeini}},\ }\bibfield  {title}
	{\bibinfo {title} {Resolving the energy levels of a nanomechanical
			oscillator},\ }\href {https://doi.org/10.1038/s41586-019-1386-x} {\bibfield
		{journal} {\bibinfo  {journal} {Nature}\ }\textbf {\bibinfo {volume} {571}},\
		\bibinfo {pages} {537} (\bibinfo {year} {2019})}\BibitemShut {NoStop}%
	\bibitem [{\citenamefont {Bothner}\ \emph {et~al.}(2022)\citenamefont
		{Bothner}, \citenamefont {Rodrigues},\ and\ \citenamefont
		{Steele}}]{bothner_four-wave-cooling_2022}%
	\BibitemOpen
	\bibfield  {author} {\bibinfo {author} {\bibfnamefont {D.}~\bibnamefont
			{Bothner}}, \bibinfo {author} {\bibfnamefont {I.~C.}\ \bibnamefont
			{Rodrigues}},\ and\ \bibinfo {author} {\bibfnamefont {G.~A.}\ \bibnamefont
			{Steele}},\ }\bibfield  {title} {\bibinfo {title} {Four-wave-cooling to the
			single phonon level in {Kerr} optomechanics},\ }\href
	{https://doi.org/10.1038/s42005-022-00808-3} {\bibfield  {journal} {\bibinfo
			{journal} {Communications Physics}\ }\textbf {\bibinfo {volume} {5}},\
		\bibinfo {pages} {1} (\bibinfo {year} {2022})}\BibitemShut {NoStop}%
	\bibitem [{\citenamefont {Luschmann}\ \emph {et~al.}(2022)\citenamefont
		{Luschmann}, \citenamefont {Schmidt}, \citenamefont {Deppe}, \citenamefont
		{Marx}, \citenamefont {Sanchez}, \citenamefont {Gross},\ and\ \citenamefont
		{Huebl}}]{luschmann_mechanical_2022}%
	\BibitemOpen
	\bibfield  {author} {\bibinfo {author} {\bibfnamefont {T.}~\bibnamefont
			{Luschmann}}, \bibinfo {author} {\bibfnamefont {P.}~\bibnamefont {Schmidt}},
		\bibinfo {author} {\bibfnamefont {F.}~\bibnamefont {Deppe}}, \bibinfo
		{author} {\bibfnamefont {A.}~\bibnamefont {Marx}}, \bibinfo {author}
		{\bibfnamefont {A.}~\bibnamefont {Sanchez}}, \bibinfo {author} {\bibfnamefont
			{R.}~\bibnamefont {Gross}},\ and\ \bibinfo {author} {\bibfnamefont
			{H.}~\bibnamefont {Huebl}},\ }\bibfield  {title} {\bibinfo {title}
		{Mechanical frequency control in inductively coupled electromechanical
			systems},\ }\href {https://doi.org/10.1038/s41598-022-05438-x} {\bibfield
		{journal} {\bibinfo  {journal} {Scientific Reports}\ }\textbf {\bibinfo
			{volume} {12}},\ \bibinfo {pages} {1608} (\bibinfo {year}
		{2022})}\BibitemShut {NoStop}%
	\bibitem [{\citenamefont {Gambetta}\ \emph {et~al.}(2006)\citenamefont
		{Gambetta}, \citenamefont {Blais}, \citenamefont {Schuster}, \citenamefont
		{Wallraff}, \citenamefont {Frunzio}, \citenamefont {Majer}, \citenamefont
		{Devoret}, \citenamefont {Girvin},\ and\ \citenamefont
		{Schoelkopf}}]{gambetta_qubit-photon_2006}%
	\BibitemOpen
	\bibfield  {author} {\bibinfo {author} {\bibfnamefont {J.}~\bibnamefont
			{Gambetta}}, \bibinfo {author} {\bibfnamefont {A.}~\bibnamefont {Blais}},
		\bibinfo {author} {\bibfnamefont {D.~I.}\ \bibnamefont {Schuster}}, \bibinfo
		{author} {\bibfnamefont {A.}~\bibnamefont {Wallraff}}, \bibinfo {author}
		{\bibfnamefont {L.}~\bibnamefont {Frunzio}}, \bibinfo {author} {\bibfnamefont
			{J.}~\bibnamefont {Majer}}, \bibinfo {author} {\bibfnamefont {M.~H.}\
			\bibnamefont {Devoret}}, \bibinfo {author} {\bibfnamefont {S.~M.}\
			\bibnamefont {Girvin}},\ and\ \bibinfo {author} {\bibfnamefont {R.~J.}\
			\bibnamefont {Schoelkopf}},\ }\bibfield  {title} {\bibinfo {title}
		{Qubit-photon interactions in a cavity: {Measurement}-induced dephasing and
			number splitting},\ }\href {https://doi.org/10.1103/PhysRevA.74.042318}
	{\bibfield  {journal} {\bibinfo  {journal} {Physical Review A}\ }\textbf
		{\bibinfo {volume} {74}},\ \bibinfo {pages} {042318} (\bibinfo {year}
		{2006})}\BibitemShut {NoStop}%
	\bibitem [{\citenamefont {Blais}\ \emph {et~al.}(2021)\citenamefont {Blais},
		\citenamefont {Grimsmo}, \citenamefont {Girvin},\ and\ \citenamefont
		{Wallraff}}]{blais_circuit_2021}%
	\BibitemOpen
	\bibfield  {author} {\bibinfo {author} {\bibfnamefont {A.}~\bibnamefont
			{Blais}}, \bibinfo {author} {\bibfnamefont {A.~L.}\ \bibnamefont {Grimsmo}},
		\bibinfo {author} {\bibfnamefont {S.}~\bibnamefont {Girvin}},\ and\ \bibinfo
		{author} {\bibfnamefont {A.}~\bibnamefont {Wallraff}},\ }\bibfield  {title}
	{\bibinfo {title} {Circuit quantum electrodynamics},\ }\href
	{https://doi.org/10.1103/RevModPhys.93.025005} {\bibfield  {journal}
		{\bibinfo  {journal} {Reviews of Modern Physics}\ }\textbf {\bibinfo {volume}
			{93}},\ \bibinfo {pages} {025005} (\bibinfo {year} {2021})}\BibitemShut
	{NoStop}%
	\bibitem [{\citenamefont {Zhang}\ \emph {et~al.}(2005)\citenamefont {Zhang},
		\citenamefont {Wang},\ and\ \citenamefont {Sun}}]{zhang_cooling_2005}%
	\BibitemOpen
	\bibfield  {author} {\bibinfo {author} {\bibfnamefont {P.}~\bibnamefont
			{Zhang}}, \bibinfo {author} {\bibfnamefont {Y.~D.}\ \bibnamefont {Wang}},\
		and\ \bibinfo {author} {\bibfnamefont {C.~P.}\ \bibnamefont {Sun}},\
	}\bibfield  {title} {\bibinfo {title} {Cooling {Mechanism} for a
			{Nanomechanical} {Resonator} by {Periodic} {Coupling} to a {Cooper} {Pair}
			{Box}},\ }\href {https://doi.org/10.1103/PhysRevLett.95.097204} {\bibfield
		{journal} {\bibinfo  {journal} {Physical Review Letters}\ }\textbf {\bibinfo
			{volume} {95}},\ \bibinfo {pages} {097204} (\bibinfo {year}
		{2005})}\BibitemShut {NoStop}%
	\bibitem [{\citenamefont {Kounalakis}\ \emph {et~al.}(2020)\citenamefont
		{Kounalakis}, \citenamefont {Blanter},\ and\ \citenamefont
		{Steele}}]{kounalakis_flux-mediated_2020}%
	\BibitemOpen
	\bibfield  {author} {\bibinfo {author} {\bibfnamefont {M.}~\bibnamefont
			{Kounalakis}}, \bibinfo {author} {\bibfnamefont {Y.~M.}\ \bibnamefont
			{Blanter}},\ and\ \bibinfo {author} {\bibfnamefont {G.~A.}\ \bibnamefont
			{Steele}},\ }\bibfield  {title} {\bibinfo {title} {Flux-mediated
			optomechanics with a transmon qubit in the single-photon ultrastrong-coupling
			regime},\ }\href {https://doi.org/10.1103/PhysRevResearch.2.023335}
	{\bibfield  {journal} {\bibinfo  {journal} {Physical Review Research}\
		}\textbf {\bibinfo {volume} {2}},\ \bibinfo {pages} {023335} (\bibinfo {year}
		{2020})}\BibitemShut {NoStop}%
	\bibitem [{\citenamefont {Gardiner}\ and\ \citenamefont
		{Zoller}(2004)}]{gardiner_quantum_2004}%
	\BibitemOpen
	\bibfield  {author} {\bibinfo {author} {\bibfnamefont {C.~W.}\ \bibnamefont
			{Gardiner}}\ and\ \bibinfo {author} {\bibfnamefont {P.}~\bibnamefont
			{Zoller}},\ }\href {https://link.springer.com/book/9783540223016} {\emph
		{\bibinfo {title} {Quantum {Noise}}}}\ (\bibinfo {year} {2004})\BibitemShut
	{NoStop}%
	\bibitem [{\citenamefont {Lindblad}(1976)}]{lindblad_generators_1976}%
	\BibitemOpen
	\bibfield  {author} {\bibinfo {author} {\bibfnamefont {G.}~\bibnamefont
			{Lindblad}},\ }\bibfield  {title} {\bibinfo {title} {On the generators of
			quantum dynamical semigroups},\ }\href {https://doi.org/10.1007/BF01608499}
	{\bibfield  {journal} {\bibinfo  {journal} {Communications in Mathematical
				Physics}\ }\textbf {\bibinfo {volume} {48}},\ \bibinfo {pages} {119}
		(\bibinfo {year} {1976})}\BibitemShut {NoStop}%
	\bibitem [{\citenamefont {Krause}\ \emph {et~al.}(2022)\citenamefont {Krause},
		\citenamefont {Dickel}, \citenamefont {Vaal}, \citenamefont {Vielmetter},
		\citenamefont {Feng}, \citenamefont {Bounds}, \citenamefont {Catelani},
		\citenamefont {Fink},\ and\ \citenamefont {Ando}}]{krause_magnetic_2022}%
	\BibitemOpen
	\bibfield  {author} {\bibinfo {author} {\bibfnamefont {J.}~\bibnamefont
			{Krause}}, \bibinfo {author} {\bibfnamefont {C.}~\bibnamefont {Dickel}},
		\bibinfo {author} {\bibfnamefont {E.}~\bibnamefont {Vaal}}, \bibinfo {author}
		{\bibfnamefont {M.}~\bibnamefont {Vielmetter}}, \bibinfo {author}
		{\bibfnamefont {J.}~\bibnamefont {Feng}}, \bibinfo {author} {\bibfnamefont
			{R.}~\bibnamefont {Bounds}}, \bibinfo {author} {\bibfnamefont
			{G.}~\bibnamefont {Catelani}}, \bibinfo {author} {\bibfnamefont {J.~M.}\
			\bibnamefont {Fink}},\ and\ \bibinfo {author} {\bibfnamefont
			{Y.}~\bibnamefont {Ando}},\ }\bibfield  {title} {\bibinfo {title} {Magnetic
			{Field} {Resilience} of {Three}-{Dimensional} {Transmons} with {Thin}-{Film}
			{AlAlO}$_\text{x}$/{Al} {Josephson} {Junctions} {Approaching} 1 {T}},\ }\href
	{https://doi.org/10.1103/PhysRevApplied.17.034032} {\bibfield  {journal}
		{\bibinfo  {journal} {Physical Review Applied}\ }\textbf {\bibinfo {volume}
			{17}},\ \bibinfo {pages} {034032} (\bibinfo {year} {2022})}\BibitemShut
	{NoStop}%
	\bibitem [{\citenamefont {Forn-Díaz}\ \emph {et~al.}(2019)\citenamefont
		{Forn-Díaz}, \citenamefont {Lamata}, \citenamefont {Rico}, \citenamefont
		{Kono},\ and\ \citenamefont {Solano}}]{forn-diaz_ultrastrong_2019}%
	\BibitemOpen
	\bibfield  {author} {\bibinfo {author} {\bibfnamefont {P.}~\bibnamefont
			{Forn-Díaz}}, \bibinfo {author} {\bibfnamefont {L.}~\bibnamefont {Lamata}},
		\bibinfo {author} {\bibfnamefont {E.}~\bibnamefont {Rico}}, \bibinfo {author}
		{\bibfnamefont {J.}~\bibnamefont {Kono}},\ and\ \bibinfo {author}
		{\bibfnamefont {E.}~\bibnamefont {Solano}},\ }\bibfield  {title} {\bibinfo
		{title} {Ultrastrong coupling regimes of light-matter interaction},\ }\href
	{https://doi.org/10.1103/RevModPhys.91.025005} {\bibfield  {journal}
		{\bibinfo  {journal} {Reviews of Modern Physics}\ }\textbf {\bibinfo {volume}
			{91}},\ \bibinfo {pages} {025005} (\bibinfo {year} {2019})}\BibitemShut
	{NoStop}%
\end{thebibliography}

%

\newpage

\appendix
\section{}

Spectrum of the cavity mode is calculated from Eq.~\ref{spectrum equation} in the main 
text. In the cavity operator, it can be written as,
\begin{equation}
S_{a}(\omega) = 
\frac{1}{2\pi} \int_{-\infty}^{+\infty} \langle (\hat{a}[\omega^\prime])^\dagger\hat{a}[\omega]\rangle \mathrm{d}\omega^\prime.
\label{cavity spectrum equation}
\end{equation}

Eq.~\ref{solution} is used to calculate the steady state value 
of $\hat{a}[\omega]$, which can be written as 
\begin{equation}
\hat{a}[\omega] = (\mathcal{B}~r[\omega])_{11} = \sum_j \mathcal{B}_{1j}(r[\omega])_{j1},
\label{close solution}
\end{equation}
where $r[\omega]$ is a column matrix of noise operators of all the modes. 
\begin{equation}
r[\omega] = \begin{bmatrix}
\sqrt{\kappa_{ex}}~\hat{a}_{in}[\omega] + \sqrt{\kappa_{0}}~\hat{f}_{in}[\omega] \\
\sqrt{\kappa_{ex}}~(\hat{a}_{in}^{\dagger})[\omega] + \sqrt{\kappa_{0}}~(\hat{f}_{in}^{\dagger})[\omega] \\
\sqrt{\Gamma_{ex}}~\hat{c}_{in}[\omega] + \sqrt{\Gamma_{0}}~\hat{\xi}_{in}[\omega] \\
\sqrt{\Gamma_{ex}}~(\hat{c}_{in}^{\dagger})[\omega] + \sqrt{\Gamma_{0}}~(\hat{\xi}_{in}^{\dagger})[\omega] \\
\sqrt{\gamma_m}~\hat{b}_{in}[\omega] \\
\sqrt{\gamma_m}~(\hat{b}_{in}^{\dagger})[\omega]
\end{bmatrix}
\end{equation}

\begin{widetext}
The noise operators in the frequency domain satisfy the following relations,
\begin{subequations}
	\begin{equation}
	\langle \hat{a}_{in}[\omega] (\hat{a}_{in}[\omega^\prime])^\dagger \rangle = 2\pi\delta(\omega - \omega^\prime);~\langle (\hat{a}_{in}[\omega])^\dagger \hat{a}_{in}[\omega^\prime]  \rangle = 0
	\end{equation}
	\begin{equation}
	\langle \hat{f}_{in}[\omega] (\hat{f}_{in}[\omega^\prime])^\dagger \rangle = 2\pi\delta(\omega - \omega^\prime);~\langle (\hat{f}_{in}[\omega])^\dagger \hat{f}_{in}[\omega^\prime]  \rangle = 0
	\end{equation}
	
	\begin{equation}
	\langle \hat{c}_{in}[\omega] (\hat{c}_{in}[\omega^\prime])^\dagger \rangle = 2\pi\delta(\omega - \omega^\prime);~\langle (\hat{c}_{in}[\omega])^\dagger \hat{c}_{in}[\omega^\prime]  \rangle = 0
	\end{equation}
	\begin{equation}
	\langle \hat{\xi}_{in}[\omega] (\hat{\xi}_{in}[\omega^\prime])^\dagger \rangle = 2\pi\delta(\omega - \omega^\prime);~\langle (\hat{\xi}_{in}[\omega])^\dagger \hat{\xi}_{in}[\omega^\prime]  \rangle = 0
	\end{equation}
	\begin{equation}
	\langle \hat{b}_{in}[\omega] (\hat{b}_{in}[\omega^\prime])^\dagger \rangle = 2\pi(n_m^i + 1)\delta(\omega - \omega^\prime)
	\end{equation}
	\begin{equation}
	\langle (\hat{b}_{in}[\omega])^\dagger \hat{b}_{in}[\omega^\prime]  \rangle = 2\pi n_m^i \delta(\omega - \omega^\prime),
	\end{equation}
	\label{noise identites}
\end{subequations}

where $n_m^i$ is the thermal phonon occupancy of the mechanical mode.
We can expand the Eq.~\ref{close solution} and write the solution of $\hat{a}[\omega]$ as,
\begin{multline}
\hat{a}[\omega] =\sqrt{\kappa_{ex}}\mathcal{B}_{11}[\omega]~\hat{a}_{in}[\omega] + \sqrt{\kappa_{0}}\mathcal{B}_{11}[\omega]~\hat{f}_{in}[\omega] +                 \sqrt{\kappa_{ex}}\mathcal{B}_{12}[\omega]~(\hat{a}_{in}^{\dagger})[\omega] \\+ \sqrt{\kappa_{0}}\mathcal{B}_{12}[\omega]~(\hat{f}_{in}^{\dagger})[\omega] +  \sqrt{\Gamma_{ex}}\mathcal{B}_{13}[\omega]~\hat{c}_{in}[\omega] + \sqrt{\Gamma_{0}}\mathcal{B}_{13}[\omega]~\hat{\xi}_{in}[\omega] \\+     \sqrt{\Gamma_{ex}}\mathcal{B}_{14}[\omega]~(\hat{c}_{in}^{\dagger})[\omega] + \sqrt{\Gamma_{0}}\mathcal{B}_{14}[\omega]~(\hat{\xi}_{in}^{\dagger})[\omega] +               \sqrt{\gamma_m}\mathcal{B}_{15}[\omega]~\hat{b}_{in}[\omega] \\+ \sqrt{\gamma_m}\mathcal{B}_{16}[\omega]~(\hat{b}_{in}^{\dagger})[\omega].
\end{multline}
By using the identity $(x^\dagger)[\omega] = (x[-\omega])^\dagger $, we can re-write the solution of $\hat{a}[\omega]$. 
\begin{multline}
\hat{a}[\omega] =\sqrt{\kappa_{ex}}\mathcal{B}_{11}[\omega]~\hat{a}_{in}[\omega] + \sqrt{\kappa_{0}}\mathcal{B}_{11}[\omega]~\hat{f}_{in}[\omega] +                 \sqrt{\kappa_{ex}}\mathcal{B}_{12}[\omega]~(\hat{a}_{in}[-\omega])^{\dagger} \\+ \sqrt{\kappa_{0}}\mathcal{B}_{12}[\omega]~(\hat{f}_{in}[-\omega])^{\dagger} +  \sqrt{\Gamma_{ex}}\mathcal{B}_{13}[\omega]~\hat{c}_{in}[\omega] + \sqrt{\Gamma_{0}}\mathcal{B}_{13}[\omega]~\hat{\xi}_{in}[\omega] \\+     \sqrt{\Gamma_{ex}}\mathcal{B}_{14}[\omega]~(\hat{c}_{in}[-\omega])^{\dagger} + \sqrt{\Gamma_{0}}\mathcal{B}_{14}[\omega]~(\hat{\xi}_{in}[-\omega])^{\dagger} +               \sqrt{\gamma_m}\mathcal{B}_{15}[\omega]~\hat{b}_{in}[\omega] \\+ \sqrt{\gamma_m}\mathcal{B}_{16}[\omega]~(\hat{b}_{in}[-\omega])^{\dagger}.
\end{multline}
From the above equation and Eq.~\ref{noise identites}, we 
can calculate $\langle (\hat{a}[\omega^\prime])^\dagger\hat{a}[\omega]\rangle$,

\begin{multline}
\langle (\hat{a}[\omega^\prime])^\dagger\hat{a}[\omega]\rangle = 2\pi\kappa_{ex}\mathcal{B}_{12}^{\ast}[\omega^\prime]\mathcal{B}_{12}[\omega]~\delta(\omega - \omega^\prime) \\+ 
2\pi\kappa_{0}\mathcal{B}_{12}^{\ast}[\omega^\prime]\mathcal{B}_{12}[\omega]~\delta(\omega - \omega^\prime) \\+ 
2\pi\Gamma_{ex}\mathcal{B}_{14}^{\ast}[\omega^\prime]\mathcal{B}_{14}[\omega]~\delta(\omega - \omega^\prime) \\+ 
2\pi\Gamma_{0}\mathcal{B}_{14}^{\ast}[\omega^\prime]\mathcal{B}_{14}[\omega]~\delta(\omega - \omega^\prime) \\+ 
2\pi n_m^i \gamma_m\mathcal{B}_{15}^{\ast}[\omega^\prime]\mathcal{B}_{15}[\omega]~\delta(\omega^\prime - \omega) \\+ 
2\pi(n_m^i + 1)\gamma_m\mathcal{B}_{16}^{\ast}[\omega^\prime]\mathcal{B}_{16}[\omega]~\delta(\omega - \omega^\prime).
\label{correlator}
\end{multline}

Substituting this to Eq.~\ref{spectrum equation}, the spectrum 
of the cavity mode can be written as,

\begin{equation}
S_{a}(\omega) = 
n_m^i \gamma_m(|\mathcal{B}_{15}[\omega]|^2 + |\mathcal{B}_{16}[\omega]|^2) +
\kappa |\mathcal{B}_{12}[\omega]|^2 + 
\Gamma |\mathcal{B}_{14}[\omega]|^2 + 
\gamma_m|\mathcal{B}_{16}[\omega]|^2,
\label{final equation}
\end{equation}

where $\kappa$, $\Gamma$ and $\gamma_m$ are total dissipation rates of the cavity, qubit and mechanical mode respectively. $n_m^i$ is the initial mechanical mode occupancy. The terms $\mathcal{B}_{12}[\omega]$, $\mathcal{B}_{14}[\omega]$, $\mathcal{B}_{15}[\omega]$, $\mathcal{B}_{16}[\omega]$ are calculated using  \textit{Wolfram Mathematica}.
\begin{subequations}
	\begin{multline}
	\resizebox{0.98\hsize}{!}{$\mathcal{B}_{12}[\omega] = -\frac{J^2 \chi_c \tilde{\chi}_c (-{i} \tilde{\chi}_q + \chi_q ({i} + 4 \eta \tilde{\chi}_q))}
		{-4 {i} \eta^2 \chi_q \tilde{\chi}_q + J^2 (\chi_c - \tilde{\chi}_c) (-{i} \tilde{\chi}_q + \chi_q ({i} + 4 \eta \tilde{\chi}_q)) + 
			{i} (1 + g^2 (\chi_q - \tilde{\chi}_q) (\chi_m - \tilde{\chi}_m)) + 4 g^2 \eta \chi_q \tilde{\chi}_q (\chi_m - \tilde{\chi}_m)}$}
	\end{multline}
	\begin{multline}
	\resizebox{0.98\hsize}{!}{$\mathcal{B}_{14}[\omega] = \frac{J \chi_c (1 + 2 {i} \eta \chi_q) \tilde{\chi}_q}{-4 {i} \eta^2 \chi_q \tilde{\chi}_q + 
			J^2 (\chi_c - \tilde{\chi}_c) (-{i} \tilde{\chi}_q + \chi_q ({i} + 4 \eta \tilde{\chi}_q)) + 
			{i} (1 + g^2 (\chi_q - \tilde{\chi}_q) (\chi_m - \tilde{\chi}_m)) + 4 g^2 \eta \chi_q \tilde{\chi}_q (\chi_m - \tilde{\chi}_m)}$}
	\end{multline}
	
	\begin{multline}
	\resizebox{0.98\hsize}{!}{$\mathcal{B}_{15}[\omega] = -\frac{g J \chi_c (-{i} \tilde{\chi}_q + \chi_q ({i} + 4 \eta \tilde{\chi}_q)) \chi_m}{-4 {i} \eta^2 \chi_q \tilde{\chi}_q + 
			J^2 (\chi_c - \tilde{\chi}_c) (-{i} \tilde{\chi}_q + \chi_q ({i} + 4 \eta \tilde{\chi}_q)) + 
			{i} (1 + g^2 (\chi_q - \tilde{\chi}_q) (\chi_m - \tilde{\chi}_m)) + 4 g^2 \eta \chi_q \tilde{\chi}_q (\chi_m - \tilde{\chi}_m)}$}
	\label{b15 equation}
	\end{multline}
	\begin{multline}
	\resizebox{0.98\hsize}{!}{$\mathcal{B}_{16}[\omega] = -\frac{g J \chi_c (-{i} \tilde{\chi}_q + \chi_q ({i} + 4 \eta \tilde{\chi}_q)) \tilde{\chi}_m}{-4 {i} \eta^2 \chi_q \tilde{\chi}_q + 
			J^2 (\chi_c - \tilde{\chi}_c) (-{i} \tilde{\chi}_q + \chi_q ({i} + 4 \eta \tilde{\chi}_q)) + 
			{i} (1 + g^2 (\chi_q - \tilde{\chi}_q) (\chi_m - \tilde{\chi}_m)) + 4 g^2 \eta \chi_q \tilde{\chi}_q (\chi_m - \tilde{\chi}_m)}$}.
	\end{multline}
\end{subequations}

\end{widetext}

\maketitle
~
\newpage
~
\newpage

\section{}

From the Lindblad formalism the time-domain master equation of the density operator $\dot{\hat{\rho}}(t)$ is written as,
\begin{widetext}

\begin{multline}
\dot{\hat{\rho}} = {i}[\hat{\rho}, \tilde{\mathcal{H}}] + \kappa (n_c^i +1)\mathcal{D}[\hat{a}]\hat{\rho} + \kappa n_c^i \mathcal{D}[\hat{a}^\dagger]\hat{\rho} + \Gamma (n_q^i +1)\mathcal{D}[\hat{c}]\hat{\rho} + \Gamma n_q^i \mathcal{D}[\hat{c}^\dagger]\hat{\rho} + \frac{\Gamma_{\phi}}{2}\mathcal{D}[\hat{c}^\dagger\hat{c}]\hat{\rho} \\+ \gamma_m (n_m^i +1)\mathcal{D}[\hat{b}]\hat{\rho} + \gamma_m n_m^i \mathcal{D}[\hat{b}^\dagger]\hat{\rho}.
\end{multline}

\end{widetext}

Here $\kappa$ and $\gamma_m$ are energy relaxation rates of cavity and mechanical mode. Qubit relaxation and pure dephasing are represented as $\Gamma$ and $\Gamma_{\phi}$. The initial thermal occupancy of the cavity, qubit and the mechanical modes are $n_c^i$, $n_q^i$, and $n_m^i$ respectively. For our calculation we have considered $\Gamma_{\phi} = 0$. $\mathcal{D}[\hat{\mathcal{O}}]$ is the Lindblad super-operator written as,
\begin{equation}
\mathcal{D}[\hat{\mathcal{O}}]\hat{\rho} := \hat{\mathcal{O}} \hat{\rho} \hat{\mathcal{O}}^\dagger - \frac{1}{2} \hat{\mathcal{O}}^\dagger \hat{\mathcal{O}}\hat{\rho} - \frac{1}{2} \hat{\rho}\hat{\mathcal{O}}^\dagger\hat{\mathcal{O}}.
\end{equation}
We write down the equation of motion from the Hamiltonian in Eq.~(5). This is to calculate expectation values of different operators. The coupled linear equations are written in the matrix form,
\begin{equation}
\dot{\boldsymbol{d}} = \mathcal{M}\boldsymbol{d} + \mathcal{N},
\end{equation}
where $\boldsymbol{d}$ is the column matrix consisting of the expectations values.

\begin{widetext}
	
\begin{equation}
\boldsymbol{d} = 
\begin{bmatrix}
\langle\hat{a}^{\dagger} \hat{a}\rangle \\
\langle\hat{b}^{\dagger} \hat{b}\rangle \\
\langle\hat{c}^{\dagger} \hat{c}\rangle \\
\langle\hat{a}^2\rangle \\
\langle\hat{a}^{\dagger 2}\rangle \\
\langle\hat{b}^2\rangle \\
\langle\hat{b}^{\dagger 2}\rangle \\
\langle\hat{c}\rangle \\
\langle\hat{c}^{\dagger 2}\rangle \\
\langle\hat{a} \hat{b}\rangle \\
\langle\hat{a}^{\dagger} \hat{b}^{\dagger}\rangle \\
\langle\hat{a}^{\dagger} \hat{b}\rangle \\
\langle\hat{a} \hat{b}^{\dagger}\rangle \\
\langle\hat{c} \hat{b}\rangle \\
\langle\hat{c}^{\dagger} \hat{b}^{\dagger}\rangle \\
\langle\hat{c}^{\dagger} \hat{b}\rangle \\
\langle\hat{c} \hat{b}^{\dagger}\rangle \\
\langle\hat{a} \hat{c}\rangle \\
\langle\hat{a}^{\dagger} \hat{c}^{\dagger}\rangle \\
\langle\hat{a}^{\dagger} \hat{c}\rangle \\
\langle\hat{a} \hat{c}^{\dagger}\rangle \\
\end{bmatrix}
\end{equation}

\begin{equation}
\mathcal{M} = 
\resizebox{.8\textwidth}{!}{$\begin{bmatrix}
	-\kappa & 0 & 0 & 0 & 0 & 0 & 0 & 0 & 0 & 0 & 0 & 0 & 0 & 0 & 0 & 0 & 0 & {i}J & -{i}J & -{i}J & {i}J \\
	0 & -\gamma_m & 0 & 0 & 0 & 0 & 0 & 0 & 0 & 0 & 0 & 0 & 0 & {i}g & -{i}g & {i}g & -{i}g & 0 & 0 & 0 & 0 \\
	0 & 0 & -\Gamma & 0 & 0 & 0 & 0 & -2{i}\eta & 2{i}\eta & 0 & 0 & 0 & 0 & {i}g & -{i}g & -{i}g & {i}g & {i}J & -{i}J & {i}J & -{i}J \\
	0 & 0 & 0 & 1/\chi_{\hat{a}\hat{a}} & 0 & 0 & 0 & 0 & 0 & 0 & 0 & 0 & 0 & 0 & 0 & 0 & 0 & -2{i}J & 0 & 0 & -2{i}J \\
	0 & 0 & 0 & 0 & 1/\chi_{\hat{a}\hat{a}}^{\ast} & 0 & 0 & 0 & 0 & 0 & 0 & 0 & 0 & 0 & 0 & 0 & 0 & 0 & 2{i}J & 2{i}J & 0 \\
	0 & 0 & 0 & 0 & 0 & 1/\chi_{\hat{b}\hat{b}} & 0 & 0 & 0 & 0 & 0 & 0 & 0 & -2{i}g & 0 & -2{i}g & 0 & 0 & 0 & 0 & 0 \\
	0 & 0 & 0 & 0 & 0 & 0 & 1/\chi_{\hat{b}\hat{b}}^{\ast} & 0 & 0 & 0 & 0 & 0 & 0 & 0 & 2{i}g & 0 & 2{i}g & 0 & 0 & 0 & 0 \\
	0 & 0 & 4{i}\eta & 0 & 0 & 0 & 0 & 1/\chi_{\hat{c}\hat{c}} & 0 & 0 & 0 & 0 & 0 & -2{i}g & 0 & 0 & -2{i}g & -2{i}J & 0 & -2{i}J & 0 \\
	0 & 0 & -4{i}\eta & 0 & 0 & 0 & 0 & 0 & 1/\chi_{\hat{c}\hat{c}}^{\ast} & 0 & 0 & 0 & 0 & 0 & 2{i}g & 2{i}g & 0 & 0 & 2{i}J & 0 & 2{i}J \\
	0 & 0 & 0 & 0 & 0 & 0 & 0 & 0 & 0 & 1/\chi_{\hat{a}\hat{b}} & 0 & 0 & 0 & -{i}J & 0 & -{i}J & 0 & -{i}g & 0 & 0 & -{i}g \\
	0 & 0 & 0 & 0 & 0 & 0 & 0 & 0 & 0 & 0 & 1/\chi_{\hat{a}\hat{b}}^{\ast} & 0 & 0 & 0 & {i}J & 0 & {i}J & 0 & {i}g & {i}g & 0 \\
	0 & 0 & 0 & 0 & 0 & 0 & 0 & 0 & 0 & 0 & 0 & 1/\chi_{\hat{a}^{\dagger}\hat{b}} & 0 & {i}J & 0 & {i}J & 0 & 0 & -{i}g & -{i}g & 0 \\
	0 & 0 & 0 & 0 & 0 & 0 & 0 & 0 & 0 & 0 & 0 & 0 & 1/\chi_{\hat{a}^{\dagger}\hat{b}}^{\ast} & 0 & -{i}J & 0 & -{i}J & {i}g & 0 & 0 & {i}g \\
	0 & -{i}g & -{i}g &     0 &     0 & -{i}g &     0 & -{i}g &     0 & -{i}J &     0 & -{i}J &     0 & 1/\chi_{\hat{c}\hat{b}} & 0 & 2{i}\eta & 0 & 0 & 0 & 0 & 0 \\
	0 &  {i}g &  {i}g &     0 &     0 &     0 &  {i}g &     0 &  {i}g &     0 &  {i}J &     0 &  {i}J & 0 & 1/\chi_{\hat{c}\hat{b}}^{\ast} & 0 & -2{i}\eta & 0 & 0 & 0 & 0 \\
	0 &  {i}g & -{i}g &     0 &     0 &  {i}g &     0 &     0 & -{i}g &  {i}J &     0 &  {i}J &     0 & -2{i}\eta & 0 & 1/\chi_{\hat{c}^{\dagger}\hat{b}} & 0 & 0 & 0 & 0 & 0 \\
	0 & -{i}g &  {i}g &     0 &     0 &     0 & -{i}g &  {i}g &     0 &     0 & -{i}J &     0 & -{i}J & 0 & 2{i}\eta & 0 & 1/\chi_{\hat{c}^{\dagger}\hat{b}}^{\ast} & 0 & 0 & 0 & 0 \\
	-{i}J &     0 & -{i}J & -{i}J &     0 &     0 &     0 & -{i}J &     0 & -{i}g &     0 &     0 & -{i}g & 0 & 0 & 0 & 0 & 1/\chi_{\hat{a}\hat{c}} & 0 & 0 & 2{i}\eta \\
	{i}J &     0 &  {i}J &     0 &  {i}J &     0 &     0 &     0 &  {i}J &     0 &  {i}g &  {i}g &     0 & 0 & 0 & 0 & 0 & 0 & 1/\chi_{\hat{a}\hat{c}}^{\ast} & -2{i}\eta & 0 \\
	-{i}J &     0 &  {i}J &     0 & -{i}J &     0 &     0 &  {i}J &     0 &     0 & -{i}g & -{i}g &     0 & 0 & 0 & 0 & 0 & 0 & 2{i}\eta & 1/\chi_{\hat{a}^{\dagger}\hat{c}} & 0 \\
	{i}J &     0 & -{i}J &  {i}J &     0 &     0 &     0 &     0 & -{i}J &  {i}g &     0 &     0 &  {i}g & 0 & 0 & 0 & 0 & -2{i}\eta & 0 & 0 & 1/\chi_{\hat{a}^{\dagger}\hat{c}}^{\ast} \\
	\end{bmatrix}$}
\end{equation}

\end{widetext}

Various susceptibilities are defined below.
\begin{subequations}
\begin{equation}
    \chi_{\hat{a}\hat{a}} = \frac{1}{2{i}\Delta_c - \kappa }
\end{equation}

\begin{equation}
    \chi_{\hat{b}\hat{b}} = \frac{1}{-2{i}\omega_m - \gamma_m}
\end{equation}

\begin{equation}
    \chi_{\hat{c}\hat{c}} = \frac{1}{2{i}\Delta_q - \Gamma - \Gamma_{\phi}}
\end{equation}

\begin{equation}
    \chi_{\hat{a}\hat{b}} = \frac{1}{{i}(\Delta_c - \omega_m) - \frac{1}{2} (\kappa + \gamma_m)}
\end{equation}

\begin{equation}
    \chi_{\hat{a}^{\dagger}\hat{b}} = \frac{1}{-{i}(\Delta_c + \omega_m) - \frac{1}{2} (\kappa + \gamma_m)}
\end{equation}

\begin{equation}
    \chi_{\hat{c}\hat{b}} = \frac{1}{{i}(\Delta_q - \omega_m) - \frac{1}{2} (\Gamma + \frac{\Gamma_{\phi}}{2} + \gamma_m)}
\end{equation}

\begin{equation}
    \chi_{\hat{c}^{\dagger}\hat{b}} = \frac{1}{-{i}(\Delta_q + \omega_m) - \frac{1}{2} (\Gamma + \frac{\Gamma_{\phi}}{2} + \gamma_m)}
\end{equation}

\begin{equation}
    \chi_{\hat{a}\hat{c}} = \frac{1}{{i}(\Delta_c + \Delta_q) - \frac{1}{2} (\Gamma + \frac{\Gamma_{\phi}}{2} + \kappa)}
\end{equation}

\begin{equation}
    \chi_{\hat{a}^{\dagger}\hat{c}} = \frac{1}{-{i}(\Delta_c - \Delta_q) - \frac{1}{2} (\Gamma + \frac{\Gamma_{\phi}}{2} + \kappa)}
\end{equation}
\end{subequations}

\begin{equation}
\mathcal{N} = 
\begin{bmatrix}
n_c^i \kappa \\
n_m^i \gamma_m \\
n_q^i \Gamma \\
0 \\
0 \\
0 \\
0 \\
2{i}\eta \\
-2{i}\eta \\
0 \\
0 \\
0 \\
0 \\
-{i}g \\
{i}g \\
0 \\
0 \\
-{i}J \\
{i}J \\
0 \\
0
\end{bmatrix}
\end{equation}

The steady-state solution of $\boldsymbol{d}$ matrix can be written as, 
\begin{equation}
\boldsymbol{d} = -\mathcal{M}^{-1}\mathcal{N}.
\label{steady-state solution}
\end{equation}
From Eq.~\ref{steady-state solution} we have calculated the final mechanical occupation $n_f$ as a function of the device parameters. The plot of $n_f$ as a function of coupling $g$ and detuning ($\tilde{\Delta}_q$) is shown in Fig.~\ref{fig4}d.

\end{document}